\documentclass[preprint,authoryear,3p]{elseart}

\usepackage[utf8]{inputenc}

\usepackage{graphicx}
\usepackage{bm}
\usepackage{hyperref}

\usepackage{empheq}
\usepackage{xspace}
\usepackage{color}
\usepackage{amssymb}
\usepackage{booktabs}
\usepackage{appendix}
\usepackage{subcaption}
\usepackage{transparent}

\hypersetup{
bookmarksnumbered = true,
unicode=false,          
pdftoolbar=true,        
pdfmenubar=true,        
pdffitwindow=false,     
pdfstartview={Fit},    
pdfauthor={P.Recho},     
pdfsubject={Subject},   
pdfcreator={Creator},   
pdfproducer={Producer}, 
pdfkeywords={keyword1} {key2} {key3}, 
pdfnewwindow=true,      
colorlinks=true,       
linkcolor=black,          
citecolor=black,        
filecolor=black,      
urlcolor=black,           
pdfdisplaydoctitle = true
}

\usepackage[normalem]{ulem}
\usepackage{color}
\usepackage{xcolor}
\colorlet{red}{black}

\journal{}

\usepackage[normalem]{ulem}
\usepackage{mathtools}
\usepackage{amsmath}
\usepackage{placeins}

\begin{document}
\text

\title{Cell crawling on a compliant substrate:\\
a biphasic relation with linear friction}

\author[liphy]{H. Chelly}
\author[liphy]{A. Jahangiri}
\author[liphy]{M. Mireux}
\author[liphy]{J. {\'E}tienne}
\ead{jocelyn.etienne@univ-grenoble-alpes.fr}
\author[oslo]{D.K. Dysthe}
\ead{d.k.dysthe@fys.uio.no}
\author[liphy]{C. Verdier}
\ead{claude.verdier@univ-grenoble-alpes.fr}
\author[liphy]{P. Recho}
\ead{pierre.recho@univ-grenoble-alpes.fr}
\address[liphy]{Universit\'e Grenoble Alpes, Laboratoire Interdisciplinaire de Physique, CNRS, F-38000 Grenoble, France}
\address[oslo]{University of Oslo, Department of Physics, Oslo, Norway}

\begin{abstract}

A living cell actively generates traction forces on its environment with its actin cytoskeleton. These forces deform the cell elastic substrate which, in turn, affects the traction forces exerted by the cell and can consequently modify the cell dynamics. By considering a cell constrained to move along a one-dimensional thin track, we take advantage of the problem geometry to explicitly derive the effective law that describes the non-local frictional contact between the cell and the deformable substrate. We then couple such a law with one of the simplest model of the active flow within the cell cytoskeleton. This offers a paradigm that does not invoke any local non-linear friction law to explain that the relation between the cell steady state velocity and the substrate elasticity is non linear as experimentally observed. Additionally, we present an experimental platform to test our theoretical predictions. While our efforts are still not conclusive in this respect as more cell types need to be investigated, our analysis of the coupling between the substrate displacement and the actin flow leads to friction coefficient estimates that are in-line with some previously reported results.

\end{abstract}

 
 \maketitle

\section{Introduction}
\begin{figure}[!ht]
  \centering
  \includegraphics[width=.5\textwidth]{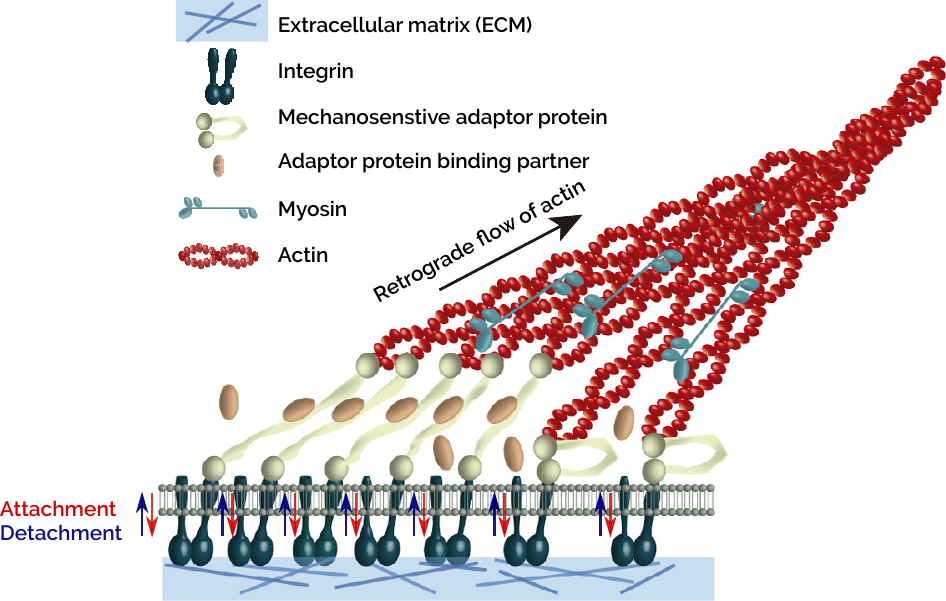}
  \caption{Composition of a focal adhesion (adapted from \cite{EloseguiArtola2018}). The force transmission from the cell to the substrate is ensured by the transmembrane proteins, integrins, connecting the actin cytoskeleton to protein receptors (e.g. fibronectin) on the substrate. The generated traction force triggers a positive feedback loop, by inducing an increase of integrins recruitement into the FA, which in return increases the traction.
  The actin retrograde flow induced by polymerization and cell contractility, is slowed down by the continuous de-/attachment of integrins to the substrate.}
  \label{contact_cell_substrate}
\end{figure}

The contact between a living cell and its substrate is mediated by focal adhesion (FAs) linking the cell cytoskeleton and its actin meshwork to the substrate. FA are constituted by a myriad of different proteins interacting with each others  in order to integrate extracellular chemical (e.g. cytokines) and mechanical (e.g. substrate rigidity, external forces) cues \citep{Geiger2009}. The cell then self-adjusts according to these cues by,  for instance, migrating from zones of lower substrate rigidity to zones of higher substrate rigidity following a process called durotaxis \citep{Lo2000}. One of the key components of this sensing machinery is the integrin, a transmembrane protein which can bind actin filaments with substrate--coating proteins, such as fibronectin. The integrin works as a bidirectional link as it is used as a mechanical sensor of the substrate but also to transmit forces  generated by the active flow of the actin meshwork. In response, the binding and unbinding of integrins creates a feedback on the actin flow. See Fig.~\ref{contact_cell_substrate}.

Minimal agent-based models of sliding friction have been explored by \cite{Srinivasan2009,Li2010,Sabass2010, Sens2013} to investigate the interaction between an actin filament and  stochastic bonds, based on the Lacker-Peskin model \citep{Lacker1986}. The results show a biphasic (i.e. bell-shape) relationship between the cell traction force and the actin velocity: at sufficiently small actin velocity, 
the loading in the bonds increases slowly,  which leads to a linear increase of the traction force with the actin velocity. For higher actin velocity, the loading rate in the bonds is high, therefore they break very rapidly and the traction force decreases with the actin velocity. Using this microscopic approach, \cite{Sens2020} formulated a simple cell model consisting of one stochastic adhesion module linked by a spring at both ends. This local non-linear dependence of the traction force on the actin velocity has been used to explain the well-established experimental observation that the global cell velocity depends on the fibronectin ligand density in a biphasic way  \cite{dimilla1991mathematical,palecek1997integrin}. Similarly, starting from the general trend that stationary cells generate large traction forces while showing a slow actin flow and on the contrary motile cells generate small traction forces with a fast retrograde flow, \cite{Barnhart2015} deduced that the frictional slippage is weaker at the rear of a motile cell than in stationary cells, and thus introduced a non-linear actin flow-dependent friction coefficient to ensure a decreased cell-substrate coupling above a critical actin flow velocity.   

Nevertheless, the experimental observation \citep{Schwarz2012} that, below a velocity threshold, the transmitted traction force evolves linearly with the actin velocity, has led to disregard this complexity and consider the simple viscous friction law $\bm{T} = \xi \bm{v}$, where $\bm{T}$ represents the traction force exerted by the cell on the substrate, $\bm{v}$ is the actin  flow velocity and $\xi$ an effective viscous friction coefficient (see \citep{Recho2015} and references therein for a detailed discussion concerning this reduction). In order to account for a non-uniform distribution of the adhesion bonds under the cell (higher traction in the vicinity of the cell boundary), a space dependent friction coefficient has  been introduced in several studies \citep{Rubinstein2009, Mogilner2003, roux2016prediction}. 

But the linear viscous friction model formulated above is only suited for infinitely rigid substrates which do not deform under cell traction forces. While this situation is standard \emph{in vitro}, substrates with a compliance that matches the one of the cell  are commonly encountered \emph{in vivo} and the traction forces exerted by the cell can induce a motion of the substrate. In order to generalize the  model to the case of a deformable substrate, a simple approach is to consider the difference between the actin velocity $\bm{v}$ and the substrate velocity $\bm{v^s}$ in the friction law, such that $\bm{T} = \xi (\bm{v}-\bm{v^s})$. \cite{Wong2011} implemented this friction law with a space-dependent friction coefficient to model the interaction between a hyperelastic cell and a hyperelastic substrate. \cite{Hassan2019} also used this law combined with a phase-field model for the cell, where a migration directional bias was introduced  in order to promote cell migration in the direction of increasing substrate rigidity. More recently, \cite{Zhang2020} used a similar approach with an actin polymerization bias towards substrate tensile regions (resulting from external forces or cell traction forces) to promote cell migration in these directions. From another perspective which does not explicitly solve force balance within the cell, \cite{Ziebert2013, Loeber2014} coupled a phase-field framework \citep{Biben2003},  with reaction-diffusion adhesion dynamics involving a substrate displacement-dependent detachment to ensure the gripping (resp. slipping) below (resp. above) a substrate displacement critical value. While \cite{Ziebert2013} used a single global spring to model the elastic substrate in the fashion of \cite{Chan2008}, \cite{Loeber2014} improved this previous model by locally resolving the viscoelastic incompressible substrate displacement. In these models a local traction force exerted by a cell generates a non-local displacement of the substrate under the cell and therefore induces a local reorganization of the cell cytoskeleton which ultimately impacts its motility dynamics in a non trivial way.

In the present paper, by extending one of the simplest model of the cytoskeleton actin turnover driving cell crawling \citep{kruse2006contractility,Juelicher2007}  to the case of a deformable elastic substrate, we show that the account of the substrate displacement  is sufficient to explain a biphasic relation relating the cell velocity to the substrate stiffness \citep{Stroka2009,Peyton2005}. Therefore it is not necessary to invoke a non-linear dependence of the traction force on the actin velocity as previously discussed or indeed a direct dependence of the active force production machinery on the substrate stiffness \citep{Dokukina2010, sarvestani2011model} to reach this conclusion. 

In order to understand the influence of the substrate rigidity on cell motility from a purely mechanical standpoint, we need a relation linking the flow of the cell cytoskeleton to substrate deformation. For simplicity we assume a semi-infinite elastic substrate, implying that the thickness of the extra-cellular matrix (ECM) substrate is large compared to the cell size as opposed to the situation studied in \citep{nicolas2004elastic} where the ECM and adhesion plane form a thin film. The two opposed limits (thick or thin ECM) have been further studied in \citep{nicolas2006limitation} in the context of the focal adhesion size regulation. In this semi-infinite case, the non-local response of the compliant substrate to a local traction force within the small deformation framework, is given by the Cerruti-Boussinesq solution in the three dimensional case. However if the geometry of the contact is invariant in one direction, a plane strain assumption can apply, leading to a Flamant problem \citep{Johnson1987}. Both strategies have been employed to characterize the contact between the cell and the substrate. The plane strain assumption has been made by \cite{Qian2008} in order to simulate the detachment process of a focal adhesion and estimate its lifetime, and by \cite{Lelidis2013} to investigate the effect of substrate rigidity on cell motility with uniformly distributed discrete focal adhesions. The plane strain assumption represents a  simplification \textcolor{red}{valid only if the substrate deformation orthogonal to the principal loading direction is negligible}. Besides, the substrate displacement found within this framework is not bounded at infinity, thus an arbitrary length needs to be introduced, above which the displacement vanishes. These issues do not arise when considering the three dimensional problem.  Traction Force Microscopy (TFM), the method to evaluate the cell traction forces from a measured substrate displacement field, requires the resolution of an inverse problem based on the Boussinesq-Cerruti solution \citep{Dembo1996, Ambrosi2006, Michel2013, Sabass2008}. In the present work we propose an approach to obtain a simple relationship between the actin velocity and the substrate displacement under the simplifying hypotheses of a one-dimensional cell crawling on a semi-infinite incompressible elastic substrate. This model corresponds to the situation where the cell is constrained to move on a thin adhesion track.

The paper is organized as follows. In sec.~\ref{sec:model_formulation},  we build the mathematical model corresponding to our geometric assumptions by constructing the kernel governing the response of the semi-infinite elastic substrate to the cell traction force field. We operate under the hypotheses that the friction with the substrate is linear, and that the substrate is a linear elastic medium. Next, in sec.~\ref{sec:turnover}, we couple this model of the contact with one of the simplest models of actin turnover-based cell motility to obtain a coupled system relating the actin flow field, global cell velocity and substrate displacement. We then study traveling waves (TW) solutions of this problem in sec.~\ref{sec:TW} and give explicit solutions for some special cases of the actin turnover dynamics as well as asymptotic solutions when the substrate is infinitely hard (cell tractions negligible compared to the substrate stiffness)  or soft (cell tractions large compared to the substrate stiffness).  \textcolor{red}{
Then in sec.~\ref{sec:biphas}, we obtain numerically a biphasic behavior of the 
steady state velocity as a function of either the substrate rigidity or the cell--substrate friction coefficient.} In sec.~\ref{sec:platform}, we propose an experimental setup to validate this theory, using T24 bladder cancer cells migrating on substrates of different rigidities (5 kPa, 8 kPa and 28 kPa), confined along thin fibronectin--coated tracks. The actin cytoskeleton flow velocity and the substrate displacement are measured and projected along the cell major axis. Although we were not able to obtain \textcolor{red}{sufficiently} motile cells to interrogate our theoretical model in detail, we could estimate the friction coefficient for static cells on substrates of different rigidities. The experimental setup does contain some limitations to address in the future, including the lack of accuracy of the actin acquisition and the need to build thinner cell tracks to improve the applicability of the 1D hypothesis but the values obtained for the friction coefficient are already consistent with the literature describing cell crawling on a rigid substrate.

\section{Model of the contact between the cell and the substrate}\label{sec:model_formulation}

\begin{figure}[!ht]
\centering
\includegraphics[width=0.6\textwidth]{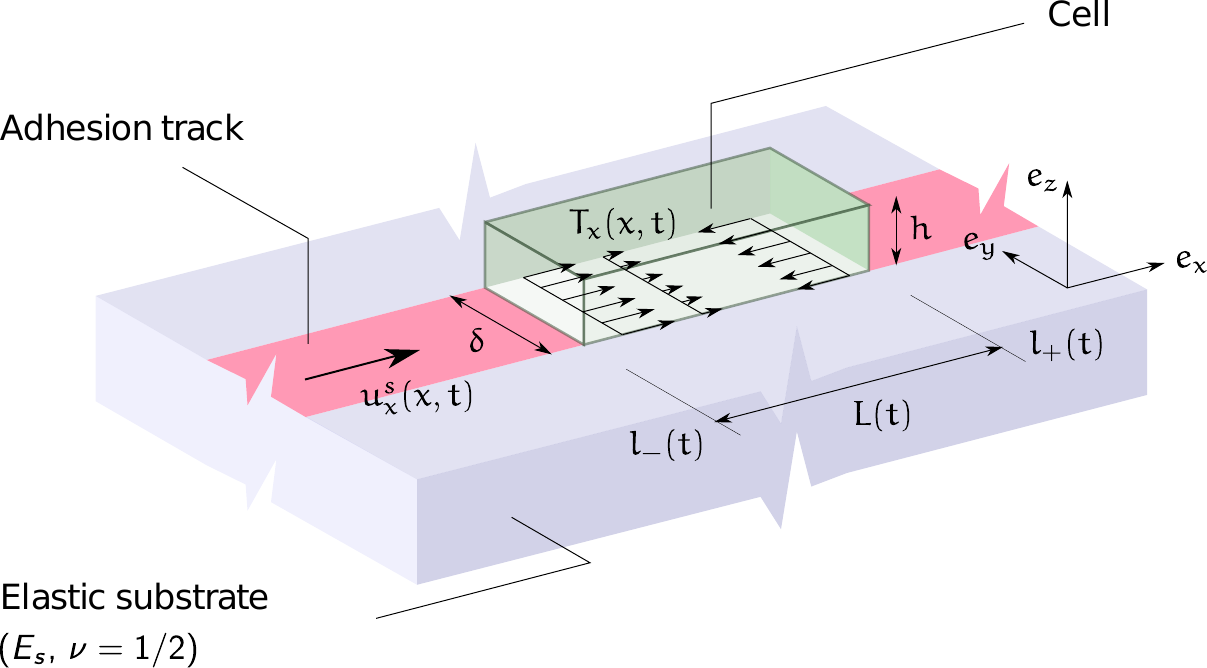}
\caption{\label{fig:scheme} Scheme of the model of a cell crawling on a semi-infinite incompressible elastic substrate. The thin adhesion track (in red) of width $\delta\ll L$ allows a one dimensional motion of the cell along the $x-$axis. The cell (in green), extending from $l_-$ to $l_+$ exerts traction forces $T_x$ on the substrate, inducing a displacement $u^s_x$ of the substrate at the surface.}
\end{figure}

We consider a cell moving on \textcolor{red}{a semi-infinite} elastic substrate and restrict our analysis to small deformations of the substrate, neglecting both physical and geometrical non-linearities of the elastic problem. In particular, we shall not differentiate the Eulerian and Lagrangian frames in our approach. The orthonormal frame of reference $(\bm{e_x},\bm{e_y},\bm{e_z})$ is chosen such that the substrate is semi-infinite \textcolor{red}{opposite to} the $\bm{e_z}$ direction and material points are labeled by the spatial coordinate $\bm{r}=(x,y,z)$.
 \textcolor{red}{We consider the case of a cell confined to move on a straight track of width $\delta$ oriented along the $x-$direction (See Fig.~\ref{fig:scheme}). The cell length $L(t)$ is defined by $L(t) = l_{+}(t)-l_{-}(t)$, where $l_{\pm}(t)$ denotes the cell fronts, and the geometric center is  $C(t)=(l_{+}(t)+l_{-}(t))/2$. Considering that the track constrains the cell so that it is very thin compared to its length ($\delta\ll L$), we assume that it only exerts traction forces tangentially to the substrate
\begin{equation}\label{eq:traction2d}
\bm{T}(\bm{r},t) = T_x(x,t)\, \bm{e_x} + T_y(x,t)\, \bm{e_y},
\end{equation}
where $x\in [l_-(t),l_+(t)]$. Taking advantage of the fact that the track is thin compared to the cell length, we have supposed that the cell fronts are straight and orthogonal to the track, and that the traction forces only depend on the $x$-coordinate along the track.}

\textcolor{red}{Defining the Boussinesq-Green kernel for a semi-infinite incompressible elastic medium of Young modulus $E_s$ \citep{Landau1959} 
$$\mathcal{G}(x,y,z)=\frac{3}{4(x^2+y^2+z^2)^{3/2}\pi E_s}\left(
\begin{array}{ccc}
 2 x^2+y^2+z^2 & x y & x z \\
 x y & x^2+2 y^2+z^2 & y z \\
 x z & y z & x^2+y^2+2 z^2 \\
\end{array}
\right),$$
 we can express the displacement of the substrate $\bm{u}$ due to the traction forces as
\begin{align}\label{eq:boussinesq}
\bm{u}(x,y,z,t) &= \int_{l_-(t)}^{l_+(t)}\int_{-\delta/2}^{\delta/2}\frac{3/(4\pi E_s)}{((x-x')^2+(y-y')^2+z^2)^{3/2}}
\nonumber\\&&\hspace*{-.35\textwidth}
\left(
\begin{array}{c}
 T_y(x',t)(x-x')(y-y')+T_x(x',t) ((2 (x-x')^2+(y-y')^2+z^2) \\
 T_x(x',t)(y-y') (x-x')+T_y(x',t) ((x-x')^2+2 (y-y')^2+z^2) \\
 (T_x(x',t) (x-x')+T_y(x',t) (y-y')) z  \\
\end{array}
\right) \mathrm dx'\mathrm dy'.&
\end{align}}

\begin{figure}[!ht]
\centering
\includegraphics[width=0.5\textwidth]{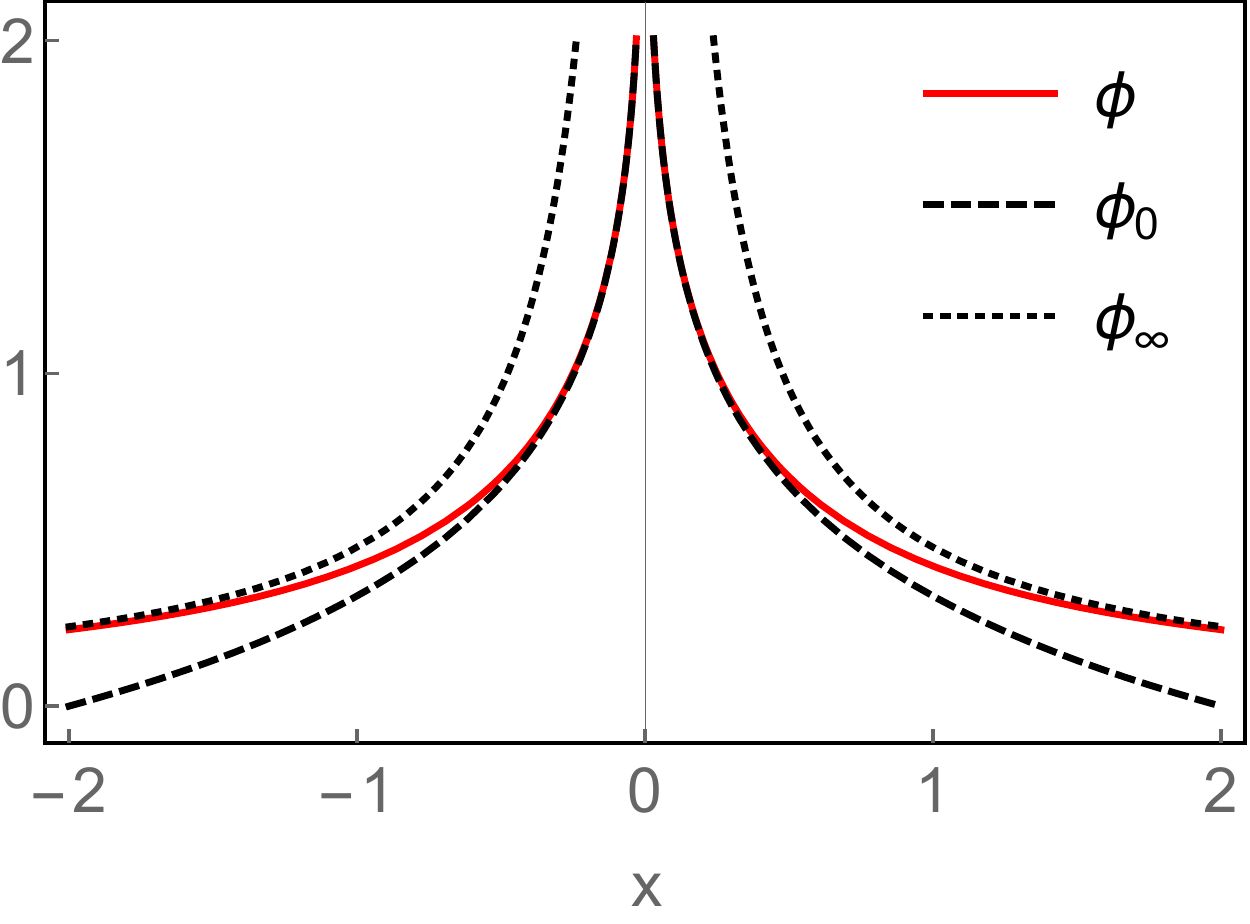}
\caption{\label{fig:kernels_compa} The non-local interaction kernel $\phi$ and the kernels $\phi_0$ and $\phi_{\infty}$ which represent the two limiting behaviors of plane strain and plane stress, respectively. Parameter $E_s=1$. }

\end{figure}

\textcolor{red}{Averaging the displacement over the $y$-direction leads to the following simplification
\begin{align}\label{eq:boussinesqtrack_1}
\bar{\bm{u}}(x,z,t) = \int_{l_-(t)}^{l_+(t)}\int_{-\delta/2}^{\delta/2}\int_{-\delta/2}^{\delta/2}\frac{3/(4\pi E_s)}{\delta((x-x')^2+(y-y')^2+z^2)^{3/2}}
\left(
\begin{array}{c}
T_x(x',t) \left(2 (x-x')^2+(y-y')^2+z^2\right) \\
T_y(x',t) \left((x-x')^2+2 (y-y')^2+z^2\right) \\
T_x(x',t)(x-x') z \\
\end{array}
\right) \mathrm dx'\mathrm dy'\mathrm dy.
\end{align}
Next, by taking the value of the displacement at $z=0$, we find  
\begin{align}\label{eq:boussinesqtrack}
\bm{u^s}(x,t) = \int_{l_-(t)}^{l_+(t)}\int_{-\delta/2}^{\delta/2}\int_{-\delta/2}^{\delta/2}\frac{3/(4\pi E_s)}{\delta((x-x')^2+(y-y')^2)^{3/2}}
\left(
\begin{array}{c}
T_x(x',t) \left(2 (x-x')^2+(y-y')^2\right) \\
T_y(x',t) \left((x-x')^2+2 (y-y')^2\right) \\
0 \\
\end{array}
\right) \mathrm dx'\mathrm dy'\mathrm dy,
\end{align}
which shows that the surface remains flat during the motion.  As we shall only investigate the force balance along the track, we project $\bm{u^s}(x,t)$ in the $x$-direction and performing the integrals in $y$, we finally obtain
\begin{equation}\label{eq:disp1D}
  u^s_x(x,t)= \int_{l_-(t)}^{l_+(t)}\phi\left(\frac{x'-x}{\delta}\right)T_x(x',t)\mathrm dx',
\end{equation}
where
\begin{equation}
  \phi(x)= \frac{3\mathrm{log}\left( \frac{1 + \sqrt{1+x^2}}{|x|}\right)}{2\pi E_s}=\frac{3\text{arcsch}\left(|x|\right)}{{2\pi E_s}}.
\end{equation}
and arcsch denotes the inverse hyperbolic cosecant.}
Note that, in the above formula, when $x\ll 1$, $\phi$ simplifies to the so-called plane strain kernel $\phi_0(x)=-3\log(|x|/2)/(2\pi E_s)$ \citep{Timoshenko1970} while it leads to the plane stress kernel $\phi_{\infty}(x)=3/(2 |x|\pi E_s)$ in the opposite limit where $x\gg 1$ \citep{Johnson1987}, see Fig.~\ref{fig:kernels_compa}. Although  $\phi_0$ is singular at $x=0$, its integral exists in the sense of Cauchy principal value.
In contrast, $\phi_{\infty}$ is singular and not integrable at $x=0$, which would lead to an infinite displacement at $x=0$. On the other hand, while $\phi_{\infty}$ tends to zero at infinity,  $\phi_0$ is unbounded at $x=\infty$, which led \cite{Timoshenko1970} to introduce an arbitrary cut-off length $x_\infty$ at which the displacement vanishes ($u^s_x(x_\infty)=0$) to regularize such situation. The kernel $\phi$ we obtain from the Boussinesq-Cerruti solution within the thin track framework, encompasses the advantages of both $\phi_0$ and $\phi_\infty$, namely the integrability at $x=0$ and the vanishing substrate displacement at $x=\infty$, because $\phi$ behaves like $\phi_0$ near the singularity $x=0$ and like $\phi_\infty$ far from the origin.

 We choose to model the traction force exerted by the cell using an effective viscous friction coefficient:
\begin{equation}\label{eq:viscous_friction}
    T_x(x,t) = \xi(v_x(x,t)-v_x^s(x,t)),
\end{equation}
where $\xi$ is an effective viscous friction coefficient that assumes a sufficiently fast attachment/detachment of the integrins, $v_x$ is the actin retrograde flow velocity and $v_x^s=\partial_t u^s_x(x,t)$ denotes the velocity of the substrate. Such relation takes into account the cell's internal activity via the actin retrograde flow, while the influence of the compliance of the substrate is encompassed in the substrate velocity. 

In order to eliminate the time-dependence of the integral boundaries, we re-map the problem using the change of variables $\bar{x}=2(x-C(t))/L(t)$ and $\bar{t}=t$, which leads to the partial derivative relations:
$$\partial_x(.)=\frac{2}{L}\partial_{\bar{x}}(.)\text{ and }\partial_t(.)\overset{\text{def}}{=}\frac{D(.)}{D\bar{t}}=\partial_{\bar{t}}(.)-\frac{2}{L}\left(\dot{C}+\frac{\dot{L}}{2}\bar{x}\right)\partial_{\bar{x}}(.),$$
where the superimposed dot denotes the time derivatives. Introducing the cell aspect ratio $\epsilon(t) = 2\delta/L(t)$ yields
\begin{equation}\label{eq:dispkernel}
  u^s_x(\bar{x},t)= \frac{L(t)}{2}\int_{-1}^{1}\phi\left(\frac{\bar{x}'-\bar{x}}{\epsilon(t)}\right)T_x(\bar{x}',t)\mathrm d\bar{x}',
\end{equation}
and injecting \eqref{eq:viscous_friction} into \eqref{eq:dispkernel} gives
\begin{equation}\label{eq:disp1D_SIDE}
  u^s_x(x,t) + \frac{\xi L(t)}{2}\int_{-1}^{1}\phi\left(\frac{x'-x}{\epsilon(t)}\right)\frac{D u^s_x(x',t)}{Dt} \mathrm dx'= \frac{\xi L(t)}{2}\int_{-1}^{1}\phi\left(\frac{x'-x}{\epsilon(t)}\right)v_x(x',t)\mathrm dx',
\end{equation}
where for simplicity of the notations and from now on, we  shall use the same symbol to denote the re-scaled variables, i.e. $\bar x := x$ and $\bar t := t$.

Assuming the actin retrograde flow velocity $v_x$ is known, \eqref{eq:disp1D_SIDE}  is a singular integro-differential equation. In the case of an infinite domain of integration, explicitly solving this type of equation is generally performed by switching to the Fourier domain. In our case where the domain of integration is a finite interval, the functional basis that diagonalizes the symmetric operator $\phi$ is not explicit as in the case of the plane strain kernel $\phi_0$  \citep{Boyd2001, Canuto2012}, making the expression of $u^s_x$ as a function of $v_x$  non-transparent. Note that while the integration of $x'$ is performed over the finite segment $[-1,1]$, $x$ takes value on the whole real line such that the boundary conditions associated to the convective time derivative $D/Dt$ are the canonical ones ($u^s_x$ and all its derivatives tend to zero \textcolor{red}{at infinity}).

\section{Coupling with a simple model of protrusion based motility}\label{sec:turnover}
One of the simplest theoretical models of cell motility that can be coupled with the traction force transmission model \eqref{eq:dispkernel} is based on the turnover of actin pushing the cell boundaries \citep{kruse2006contractility, Juelicher2007}. Modeling the cell moving along the one-dimensional track as a viscous gel with a fixed length and uniform contractile properties, we write its constitutive behavior as
\begin{equation}\label{e:const_be}
\sigma=\frac{2\eta}{L}\partial_x v_x+\chi c_0,
\end{equation}
where, $x\in[-1,1]$ is the rescaled variable introduced above, $\sigma$ is the axial stress in the cell actin meshwork, $\eta$ is the viscosity of the meshwork and $\chi c_0\geq 0$ is the uniform contraction due to the molecular motors (which we assume here to have a fixed concentration $c_0$). Additionally, denoting $h$ the cell height and assuming a thin film approximation $h\ll L$ \citep{roux2016prediction}, the force balance equation within the cell reads,
\begin{equation}\label{e:force_bal}
\frac{2 h}{L}\partial_x\sigma=\xi\left( v_x-\frac{Du^s_x(x,t)}{Dt}\right) .
\end{equation}
This equation is associated with stress boundary conditions at the cell moving edges such that 
$$\sigma\vert_{\pm 1}=\sigma_b,$$
where $\sigma_b$ is an unknown residual stress representing the constraint fixing the cell length and imposing that the two cell fronts move with the same velocity $V(t)=\dot{l}_{-}=\dot{l}_{+}$. \textcolor{red}{This boundary condition typically emerges as a limit when an effective spring that connects the two fronts has a stiffness tending to infinity while $L(t)$ tends to the rest length of the spring. The unknown residual stress $\sigma_b$ results from this double limit. See \citep{putelat2018mechanical} for more details.} Taken together with \eqref{e:const_be}, this constraint on the stress leads to 
$$\partial_xv_x\vert_{-1}=\partial_xv_x\vert_{1}.$$
Finally, the protrusion and retraction of the moving fronts are given by the Stefan boundary conditions
\begin{equation}\label{e:moving_fronts}
\dot{l}_{\pm}=v_x\vert_{\pm 1}+v_{\pm}, 
\end{equation}
where $v_{\pm}$ are the given polymerization and depolymerization velocities at the leading edge ($l_+$) and trailing edge ($l_-$). \textcolor{red}{See \cite{recho2013asymmetry} for details} We introduce the quantities $\Delta V=v_+-v_-$ representing the mismatch between polymerization and depolymerization and $V_m=(v_++v_-)/2$ representing the average turnover. More realistic albeit more complex models of actin protrusion and retraction of the cell fronts can be found in \cite{ambrosi2016mechanics,giverso2018mechanical}. \textcolor{red}{A first mechanism at the origin of the motility in the model described above is the accretion of actin at the leading edge exactly compensated by the removal of actin at the trailing edge. This leads to a propulsion at velocity $V_m$ which is independent of both the substrate and the cell constitutive behavior. Added to this mechanism, the mismatch  between polymerization and depolymerization with a preserved length $L$ of the cell implies a certain flow in the cell, exerting traction forces on the substrate proportional to $\Delta V$. This second contribution depends on the substrate stiffness as well as on the friction coefficient with the substrate in a non-trivial manner that we seek to characterize.}

We non-dimensionalize the time by $\eta/(\chi c_0)$, the distance by $L/2$ and the stress by $\chi c_0$. Combining \eqref{eq:disp1D_SIDE} with \eqref{e:const_be} and \eqref{e:force_bal}, and  denoting  $u=u_x^s$ and $v=v_x-v_0$, where 
\begin{equation}\label{e:v_0}
v_0(x)=-\Delta V\frac{\sinh(\alpha^{-1}x)}{2\sinh(\alpha^{-1})}
\end{equation} 
is a Dirichlet lift function that
accounts for the asymmetry in the protrusion/retraction kinetics at the boundaries, we obtain the non-dimensional problem 
\begin{equation}\label{e:v_turnover}
v(x,t)=\frac{1}{\alpha}\int_{-1}^{1}G\left(\frac{x-x'}{\alpha},\alpha \right) \frac{Du(x',t)}{Dt}\mathrm dx'
\qquad\text{ and }\qquad u(x,t)=\gamma\int_{-1}^{1}\Phi\left(\frac{x-x'}{\epsilon}\right)  \partial_{x'x'}[v(x',t)+v_0(x')]\mathrm dx'. 
\end{equation}
In \eqref{e:v_turnover},  $v(x,t)$ is only defined for values of $x\in[-1,1]$ (i.e. within the cell) while $u(x,t)$ is defined on the whole real line ($-\infty<x<\infty$). The three non-dimensional parameters 
$$\epsilon=\frac{2\delta}{L}\text{, }\alpha=\sqrt{\frac{4\eta h}{\xi L^2}}\text{ and }\gamma=\frac{3h\chi c_0}{\pi L E_s}.$$
respectively represent the track aspect ratio, the \textcolor{red}{ratio of the hydrodynamic length and the cell length}  and the substrate compliance \textcolor{red}{compared to the cell contractile stress}. For simplicity, we keep the same notations for the non-dimensional (de)polymerization velocities $v_{\pm}$ and their dimensional counterparts. The non-dimensional symmetric kernels representing the non-local behavior of the active viscous cytoskeleton and the elastic substrate respectively read: 
$$G\left(x,\alpha \right)=\frac{\cosh(\alpha^{-1}+x)}{2\sinh(\alpha^{-1})}-\text{H}(x)\sinh(x) 
\qquad\text{ and }\qquad
\Phi(x)=\text{arcsch}\left(|x|\right), $$
where $\text{H}$ is the Heaviside function. Thus, $G$ corresponds to the resolvent \textcolor{red}{of the elliptic problem \citep{Recho2015}
$$-\alpha^2\partial_{xx}v+v=\frac{Du}{Dt},$$
}
with periodic boundary conditions on $v$:
\begin{equation}\label{e:periodic_bc}
\partial_xv\vert_{-1}=\partial_xv\vert_{1}\text{ and } v\vert_{-1}=v\vert_1.
\end{equation}
We remark that as $\dot{L}=0$ and  $u$ is rescaled by $L/2$, the convective derivative in \eqref{e:v_turnover} takes the simple form $$\frac{Du}{Dt}=\partial_tu-V\partial_xu.$$
After $v$ and $u$ are obtained  from \eqref{e:v_turnover}, the dynamics of the moving fronts can be found from \eqref{e:moving_fronts} which leads to
\begin{equation}\label{e:front_vel}
V(t)=V_m+\frac{v\vert_{-1}+v\vert_1}{2}
\end{equation}

We provide in Table~\ref{t:valpar} some rough estimates of the various rheological coefficients entering the model. One should however bear in mind that these coefficients can vary over several orders of magnitude depending on the biological conditions and these values should therefore be taken with care. The interest of the type of reduced model that we present rather lies in capturing some physical effects with a minimal baggage than \textcolor{red}{in} describing cell crawling at a quantitative level.
\begin{table}
\begin{center}
\begin{tabular}{lll}
\hline\hline
name & symbol & typical value \\ 
\hline
cytoskeleton viscosity & $\eta$ & $100$ kPa s \citep{Juelicher2007,Rubinstein2009}\\
contractility & $\chi c_0$ &$1$ kPa \citep{Juelicher2007,Rubinstein2009} \\
 friction coefficient & $\xi$ &$10$ kPa~s~$\mu\text{m}^{-1}$ \citep{kruse2006contractility, Barnhart2011}\\
lamelipod height & $h$  &$1$ $\mu$m \citep{kruse2006contractility}\\
cell length & $L$ &$ 50$ $\mu$m \\
(de)polymerization velocities & $v_{\pm}$  &$0.2$ $\mu\text{m\,s}^{-1}$ \citep{larripa2006transport,kruse2006contractility}\\
substrate stiffness & $E_s$  & $10$ kPa\\
track width & $\delta$ & $5$  $\mu\text{m}$ \\
\hline
characteristic length & $L/2$ &$25$ $\mu$m  \\
characteristic time & $\eta/(\chi c_0)$ &  $100$ s \\
characteristic velocity & $\chi c_0 L/(2 \eta)$ & $0.25$ $\mu\text{m}\,\text{s}^{-1}$ \\
characteristic stress & $\chi c_0$ &  $1$ kPa \\
\hline
track slenderness & $\epsilon=2\delta/L$ & 0.2\\
hydrodynamic coefficient & $\alpha=\sqrt{4\eta h/(\xi L^2)}$ & 0.1 \\
substrate softness & $\gamma=3h\chi c_0/(\pi L E_s)$ & 0.02\\
turnover asymmetry & $\Delta V=2\eta (v_+-v_-)/(\chi c_0 L)$ & 1\\
turnover average & $V_m=\eta (v_++v_-)/(\chi c_0 L)$ & 1\\
\hline\hline
\end{tabular}
\end{center}
\caption{Rough estimates of the material coefficients, characteristic scales and dimensionless parameters definitions. Some parameters are subjected to several orders of magnitude variations.\label{t:valpar}}
\end{table}

\section{Traveling wave solutions}\label{sec:TW}
We now seek for traveling wave solutions of \eqref{e:v_turnover}-\eqref{e:front_vel} where $V$ is a constant and $\partial_tu=0$, implying that $Du/Dt=-V\partial_xu$. \textcolor{red}{In such a case, \eqref{e:v_turnover} can be combined into the following single integral equation:
\begin{equation}\label{e:integral_eq_TW}
\alpha^2f(x)-\gamma V\int_{-1}^1\frac{f(x')}{x'-x}\mathrm dx'-\int_{-1}^1R(x,x')f(x')\mathrm dx'=-\frac{\Delta V}{2}x+V-V_m
\end{equation}
where $f(x)=\partial_{xx}(v(x)+v_0(x))=(v(x)+v_0(x)+V\partial_x u)/\alpha^2$ represents the substrate traction force and 
$$R(x,x')=\gamma V\underset{=R_1(x,x')}{\underbrace{\frac{(x-x')/\epsilon^2}{1+(x-x')^2/\epsilon^2+\sqrt{1+(x-x')^2/\epsilon^2}}}}+\underset{=R_2(x,x')}{\underbrace{H(x-x')(x-x')+\frac{x+1}{2}x'}}$$
is a kernel that does not contain any singularity. The $R_1$ part is reminiscent of the elastic interaction with the substrate (regular part) and the $R_2$ contribution represents a second order antiderivative with appropriate boundary conditions accounted for by the righthandside of \eqref{e:integral_eq_TW}.  The unknown crawling velocity $V$ in \eqref{e:integral_eq_TW} is fixed by the global force balance constraint:
\begin{equation}\label{e:global_f_bal}
\int_{-1}^1f(x)\mathrm dx=0.
\end{equation}
}

\paragraph{Solutions for specific choices of the turnover dynamics} 

\begin{figure}[!ht]
\centering
\includegraphics[width=0.5\textwidth]{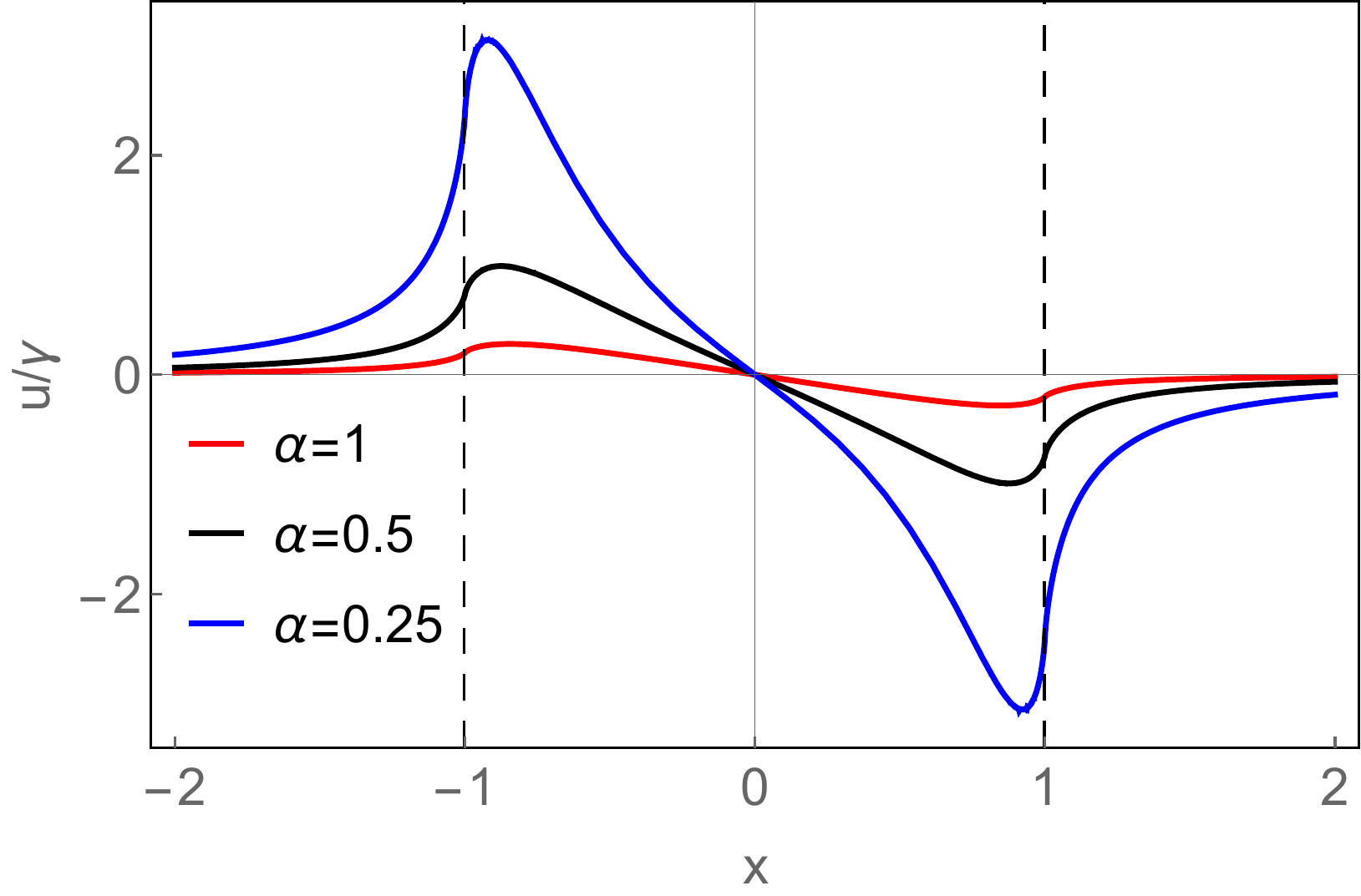}
\caption{\label{fig:sub_disp_static} Displacement $u_0$ of the substrate induced by the actin velocity field $v_x=v_0$ corresponding to a static configuration. The dashed lines indicate the normalized cell fronts. Parameters are \textcolor{red}{$\epsilon=0.2$} and $\Delta V=1$.}
\end{figure}

A simple solution \textcolor{red}{of \eqref{e:integral_eq_TW}} can be found when $v_+=-v_-$ when the two fronts are symmetrically polarized in opposite directions \textcolor{red}{leading to $V_m=0$}. \textcolor{red}{In this case, we can set $V=0$ and obtain $f(x)=f_0(x)=-\Delta V\sinh(\alpha^{-1}x)/(2\alpha^2\sinh(\alpha^{-1}))$.} This corresponds to  $v=0$ (i.e. $v_x(x) = v_0(x)$) and 
$$u(x)=\gamma u_0(x)=\gamma\int_{-1}^{1}\Phi\left(\frac{x-x'}{\epsilon}\right)  \partial_{x'x'}v_0(x')\mathrm dx'.$$
We show in Fig.\ref{fig:sub_disp_static} the typical \textcolor{red}{trend} of such substrate displacement that is induced by a traction force distribution with an even symmetry with respect to the layer center. 
The substrate displacement reaches its (anti-symmetric) maxima close to the cell boundaries and  sharply decays to zero outside of the cell as no traction forces are imposed in this region. Another simple case is when $v_+=v_-$, \textcolor{red}{leading to $f= 0$}. Thus $v=0$, $u=0$ and  $V=V_m$. This corresponds to the trivial situation where the actin protrusion and retraction are happening at the same rate at each cell edge, corresponding to a pure treadmilling movement without any internal flow of the filamentous actin. As a result, there are no traction forces and no substrate deformation. \textcolor{red}{This specific situation is rather unrealistic for most cell types as traction forces are known to be applied on the substrate during motion. } 

\textcolor{red}{More generally, we are interested in the dependence of $V$  on $\gamma$ for an arbitrary choice of $V_m$ and $\Delta V$. For this, we start by analytically analyzing the asymptotic behavior of $V$ when $\gamma\ll 1$.} 

\paragraph{Solutions in the limit of underformable substrate} When $\gamma=0$ (i.e. the substrate is undeformable), \textcolor{red}{the solution of \eqref{e:integral_eq_TW} is again $f(x)=f_0(x)$} corresponding to $v=0$, $u=0$ and $V=V_m$. This is the classical case investigated by  \cite{Juelicher2007} in the case of an infinitely stiff substrate and generalized by \cite{recho2013asymmetry} in the presence of external loading. We can further expand the solution of \eqref{e:v_turnover}-\eqref{e:front_vel} at first order when $\gamma \ll 1$
$$v(x)\simeq \gamma v_1(x)\text{ and } V\simeq V_m+\gamma V_1.$$
Note that the domain of accuracy of such approximation outside of the limit $\gamma\ll 1$ should however be investigated numerically. \textcolor{red}{The validity of such approximation clearly also relies on the fact that $V_1$ remains finite, such that $\gamma V_1$ is indeed small compared to $V_m$. }

\textcolor{red}{From \eqref{e:v_turnover}}, we obtain,
$$v_1(x)=-\frac{V_m}{\alpha}\int_{-1}^{1}G\left(\frac{x-x'}{\alpha},\alpha \right)\partial_{x'} u_0(x') \mathrm dx'$$
and thus,
$$V_1=-\frac{V_m}{\alpha}\int_{-1}^{1}G\left(\frac{1-x'}{\alpha},\alpha \right)\partial_{x'} u_0(x') \mathrm dx', $$
which after some standard manipulations takes the form:
\begin{equation}\label{e:slope}
\frac{V_1}{V_m}=-\frac{\Delta V\text{csch}^2\left(1/\alpha\right)}{4 \alpha ^4}\int_0^2 \text{arcsch}\left(\frac{x}{\epsilon }\right) \left(\alpha  \sinh \left(\frac{x}{\alpha }\right)+(x-2) \cosh \left(\frac{x}{\alpha }\right)\right)\mathrm dx.
\end{equation}

\begin{figure}[!ht]
  \begin{minipage}[b]{0.4\linewidth}
   \centering
   \includegraphics[width=7.2cm,height=4.8cm]{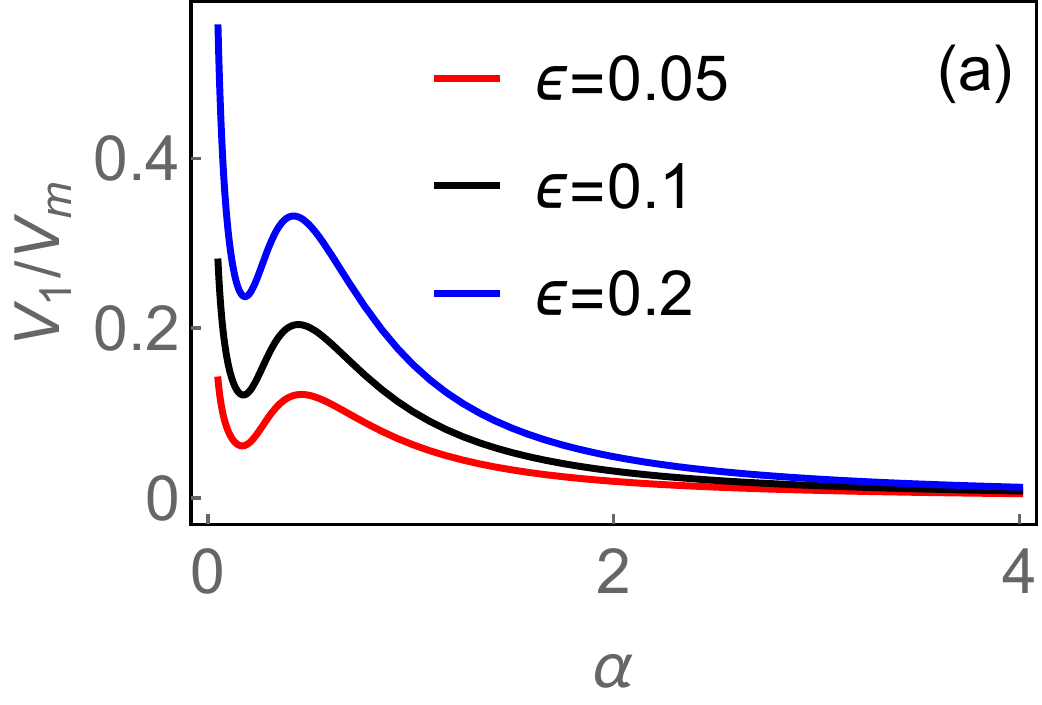}      
  \end{minipage}\hspace*{2cm}
  \begin{minipage}[b]{0.4\linewidth}
   \centering
   \includegraphics[width=7.2cm,height=4.8cm]{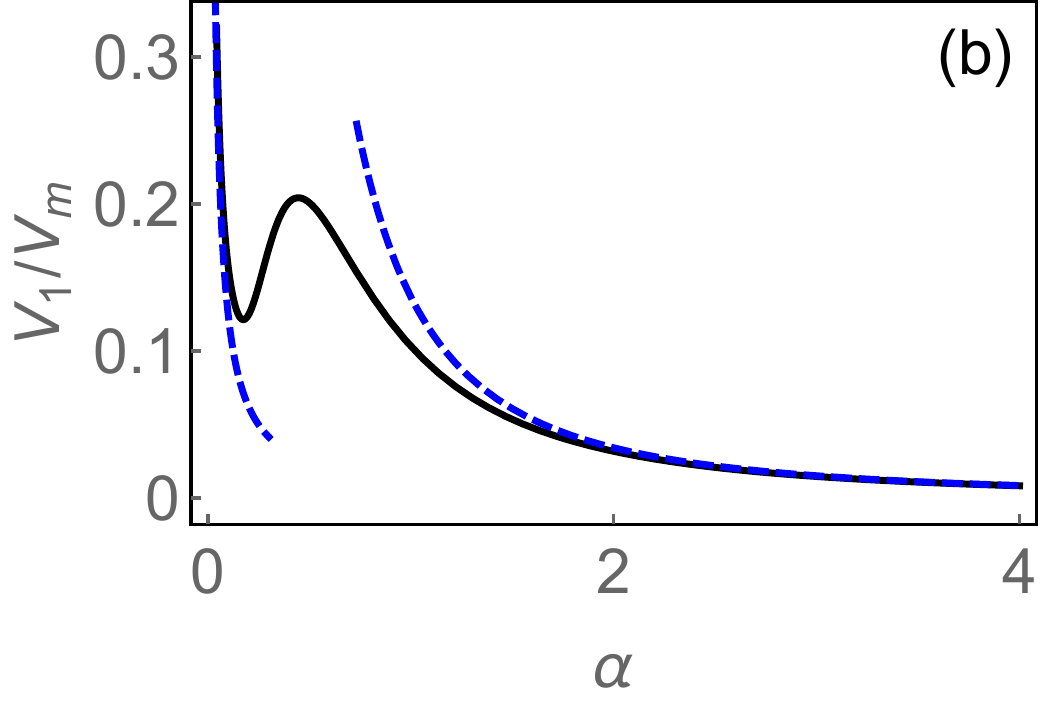}      
  \end{minipage}
  \caption{(a) Normalized value of the increase of velocity due to a small deformability of the substrate as a function of the slippage coefficient of the cell with respect to the substrate. \textcolor{red}{(b) Asymptotic regimes for large and small $\alpha$ as obtained by formulas \eqref{e:large_small_a_limits}.} Parameter $\Delta V=1$.}
  \label{fig:V_1_alpha}

\end{figure}

We illustrate the behavior of $V_1$ as a function of the parameter $\alpha$ in Fig.~\ref{fig:V_1_alpha}~(a). As $V_1$ is always positive \textcolor{red}{for a positive value of $\Delta V$}, the influence of a small substrate deformability is to increase the cell velocity compared to its value $V_m$ on an undeformable substrate \textcolor{red}{since the flow induced by $\Delta V$ in this case promotes motion in the positive direction.}  A more subtle effect is that the influence of the friction coefficient on the additional speed due to the substrate small deformability is not monotonic. In some parameter ranges, a more adhesive substrate can lead to an increased velocity since traction forces stemming from the actin flow are more effective for the propulsion. But the opposite effect is also true in other cases since an increase of adhesion can also effectively increase the friction force opposing to the motion. While this effect is lost in the limit of an undeformable substrate, we demonstrate that it can readily be observed for a slightly deformable substrate ($\gamma\ll 1$). \textcolor{red}{We further illustrate this property by analytically computing the dependence of $V_1$ in \eqref{e:slope} when $\alpha$ is either small or large to find
\begin{equation}\label{e:large_small_a_limits}
\frac{V_1}{V_m}\underset{\alpha\rightarrow 0}{\sim} \frac{\Delta V\epsilon }{4 \alpha \sqrt{\epsilon ^2+4}} \text{ and } \frac{V_1}{V_m}\underset{\alpha\rightarrow \infty}{\sim}\frac{\Delta V\epsilon  \left(2 \log \left(\frac{\sqrt{\epsilon ^2+4}+2}{\epsilon }\right)+\epsilon-\sqrt{\epsilon ^2+4} \right)}{4 \alpha^2}.
\end{equation}
These asymptotic expressions, both decreasing with $\alpha$, are plotted on Fig.~\ref{fig:V_1_alpha}~(b). The cross-over regime between them corresponds to a situation where the velocity increases with $\alpha$, in a parameter range when the corresponding hydrodynamic length is comparable to the cell length ($\alpha\sim 1$). Note that when $V_1$ blows up close to $\alpha=0$, the first order expansion breaks down as the product $\gamma V_1$ is not necessarily small any more. Thus, this behavior does not correspond to a physically meaningful regime. } This type of non-linear dependence between the cell velocity and the substrate adhesiveness was also found  in a continuum model representing an actin gel with turnover and linear friction with the substrate by \cite{CallanJones2013}. However, the origin of the effect is completely different as the authors consider an infinitely stiff substrate with an internal active stress that depends on the actin density in a non-linear fashion.

\paragraph{\textcolor{red}{Solutions in the limit of highly deformable substrate}} \textcolor{red}{We now consider the opposite limit where $\gamma\gg 1$ and the substrate is highly deformable. In this case, we seek a solution of \eqref{e:integral_eq_TW} where $f\ll 1$ while $\tilde{f}=\gamma f$ remains finite. Thus, \eqref{e:integral_eq_TW} reduces to 
\begin{equation}\label{e:int_eq_gamma_large}
- V\left( \int_{-1}^1\frac{\tilde{f}(x')}{x'-x}\mathrm dx'+\int_{-1}^1R_1(x,x')\tilde{f}(x')\mathrm dx'\right) =-\frac{\Delta V}{2}x+V-V_m
\end{equation}
Observe first that this implies that results will be independent of $\alpha$ in this limit. Indeed, any finite value of the friction coefficient has the same role in the transmission of traction forces  because the substrate is already highly deformable.
This integral equation with a singular Cauchy kernel is numerically solved following the approach of \cite{Karpenko1966approximate}.
\begin{figure}[!ht]
\centering
\includegraphics[width=0.7\textwidth]{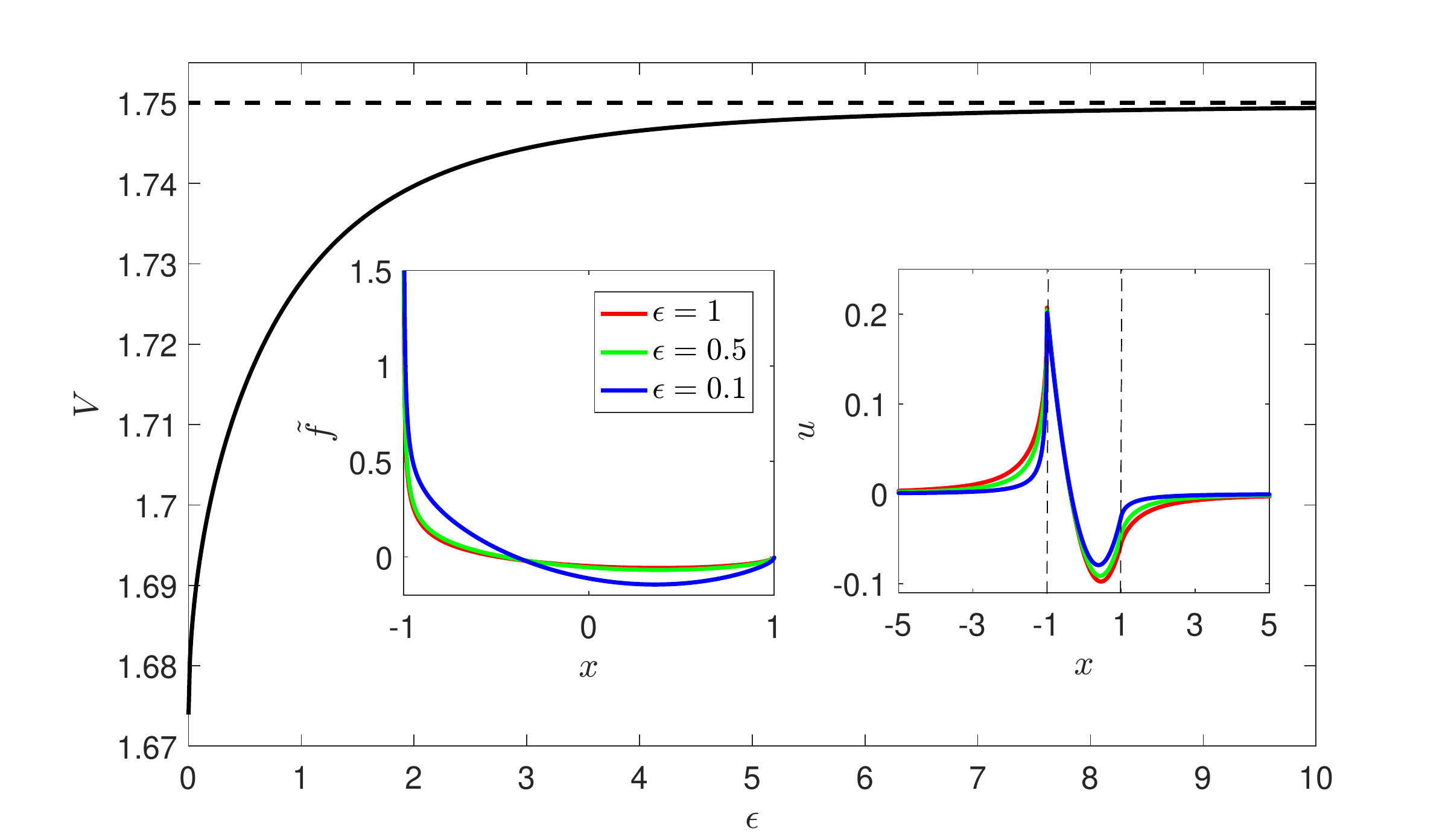}
\caption{\label{fig:V_gamma_infty} \textcolor{red}{Crawling velocity in the limit $\gamma\gg 1$ as a function of $\epsilon$. (see \eqref{e:int_eq_gamma_large}). The dashed line represents the asymptotic value $V=V_m+\Delta V/4$. In inset, we show the traction forces exerted on the substrate rescaled by $\gamma$ and the substrate displacement for some values of $\epsilon$. The traction forces display a square root singularity at the trailing edge $x=-1$. Parameters are $\Delta V=1$ and $V_m=1.5$.}}
\end{figure}
The asymptotic value of $V$ is displayed on Fig.~\ref{fig:V_gamma_infty} as a function of the track aspect ratio $\epsilon$. We also show in inset of Fig.~\ref{fig:V_gamma_infty} the spatial dependence of the traction force and the substrate displacement in this limit for several values of $\epsilon$. As expected from the combination of the two limiting relations $\partial_{xx}(v+v_0)\simeq 0$ and  $-\alpha^2\partial_{xx} v+v=-V\partial_xu$ (see \eqref{e:v_turnover}) with associated boundary conditions \eqref{e:periodic_bc}, we numerically recover that the substrate displacement on $[-1,1]$ assumes the quadratic form $u(x)=\Delta V/(4 V)x^2+(V_m/V-1)x+\text{Cst}$ where $\text{Cst}$ is a constant that remains to be set.  For large track aspect ratios (which also correspond to a plane strain situation), \eqref{e:int_eq_gamma_large} can be further simplified to 
\begin{equation}\label{e:int_eq_gamma_large_eps_large}
- V \int_{-1}^1\frac{\tilde{f}(x')}{x'-x}\mathrm dx' =-\frac{\Delta V}{2}x+V-V_m,
\end{equation}
which solution can be expressed as (see \cite{Karpenko1966approximate})
$$\tilde{f}(x)=\left( \tilde{f}_0P_0^{(1/2,-1/2)}(x)+\tilde{f}_1P_1^{(1/2,-1/2)}(x)\right) \sqrt{\frac{1-x}{1+x}}.$$
In the above expression, $P_{0,1}^{(1/2,-1/2)}$ are the first two Jacobi {polynomials} (of order zero and order one) with parameters $1/2$ and $-1/2$ and their coefficients are given by 
$$\tilde{f}_0=V-V_m-\Delta V/4\text{ and } \tilde{f}_1=-\frac{\Delta V}{2}.$$
Given the integral condition \eqref{e:global_f_bal} $\tilde{f}_0$ has to vanish, leading to the asymptotic velocity when both $\gamma$ and $\epsilon$ are large $V=V_m+\Delta V/4$.} 

\medskip

\textcolor{red}{Having clarified the two limits $\gamma\ll 1$ and $\gamma\gg 1$, which are associated to two different values of the crawling velocity and traction forces profiles, we turn to quantifying the dependence of the crawling on the substrate softness $\gamma$ and the slip coefficient with the substrate $\alpha$.}

\section{Biphasic relation of the cell velocity}\label{sec:biphas}

\textcolor{red}{Using again the approach developed in \cite{Karpenko1966approximate}, we compute numerically the solution of \eqref{e:integral_eq_TW} for the realistic parameters reported in Table~\ref{t:valpar}.  The unknown velocity in \eqref{e:integral_eq_TW} is found by dichotomy starting from the value $V=V_m$. We show on Fig.~\ref{fig:V_gamma} and Fig.~\ref{fig:V_alpha} the resulting dependence of the cell velocity driven by the internal flow as a function of the substrate softness and the slippage coefficient with the substrate.
\begin{figure}[!ht]
\centering
\includegraphics[width=0.7\textwidth]{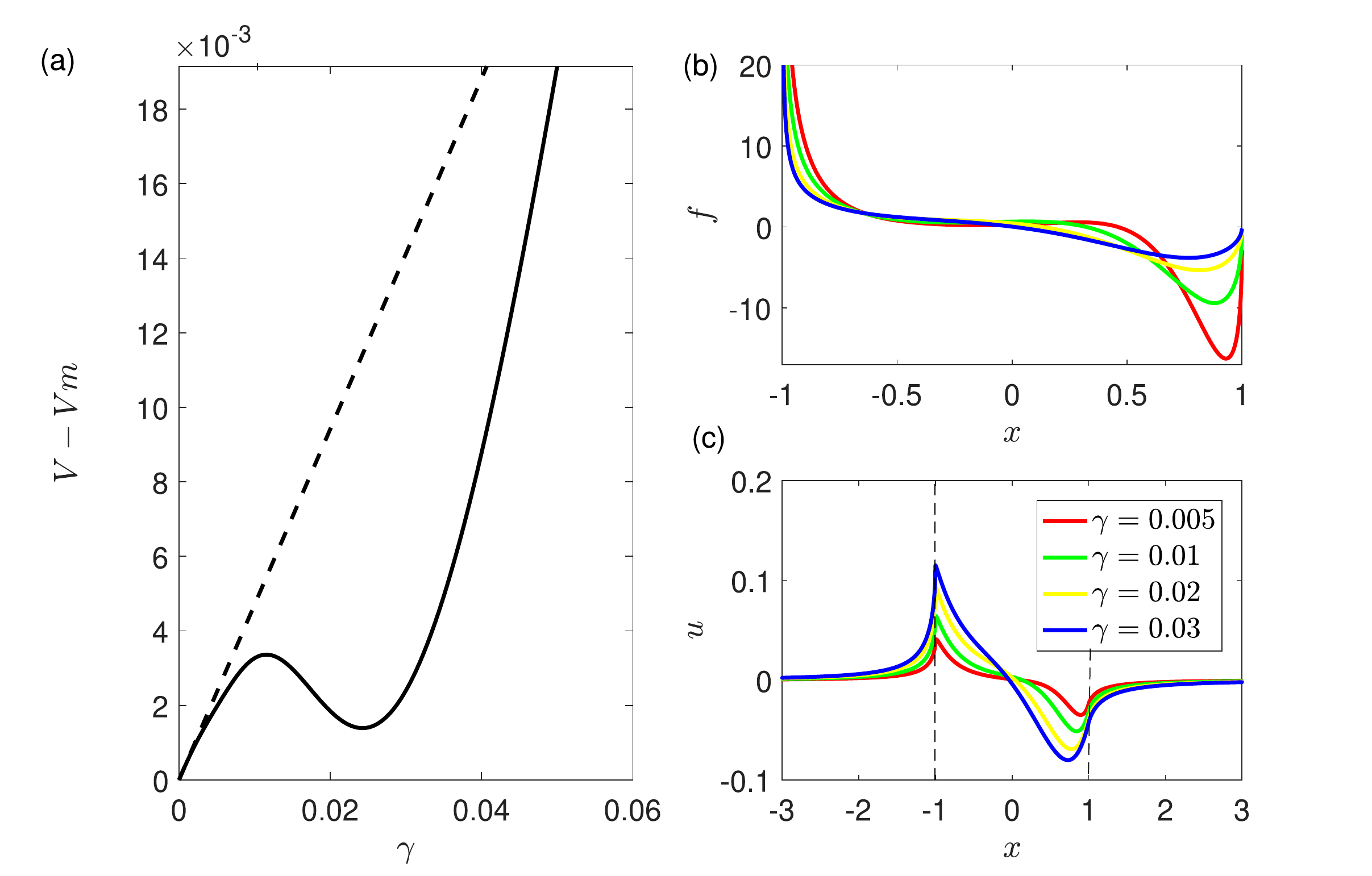}
\caption{\label{fig:V_gamma} \textcolor{red}{ Dependence of the crawling velocity of the substrate softness. (a) $V-V_m$ as a function of $\gamma$ displays a biphasic dependence in a realistic range of $\gamma$ (see Table~\ref{t:valpar}). For larger $\gamma$, the velocity increases again to reach its asymptotic value given in Fig.~\ref{fig:V_gamma_infty}. The dashed line represents the slope of the curve for small $\gamma$ given by \eqref{e:slope}. Traction forces exerted on the substrate (b) and substrate displacement (c) for several values of $\gamma$.  Parameters are $\epsilon=0.2$, $\alpha=0.1$, $\Delta V=1$ and $V_m=1.5$.}}
\end{figure}
On the left panels of these figures, we also display the typical traction force distributions exerted on the substrate and the displacement field at the contact surface as a function of the spatial coordinate. The central symmetry of these distributions is broken, which leads to a non zero flow-driven crawling velocity. The traction forces self-adjust with the substrate displacement and lead to biphasic behaviors of the crawling velocity as a function of both the substrate rigidity and adhesion, as experimentally observed in \citet{palecek1997integrin, Peyton2005}.
\begin{figure}[!ht]
\centering
\includegraphics[width=0.7\textwidth]{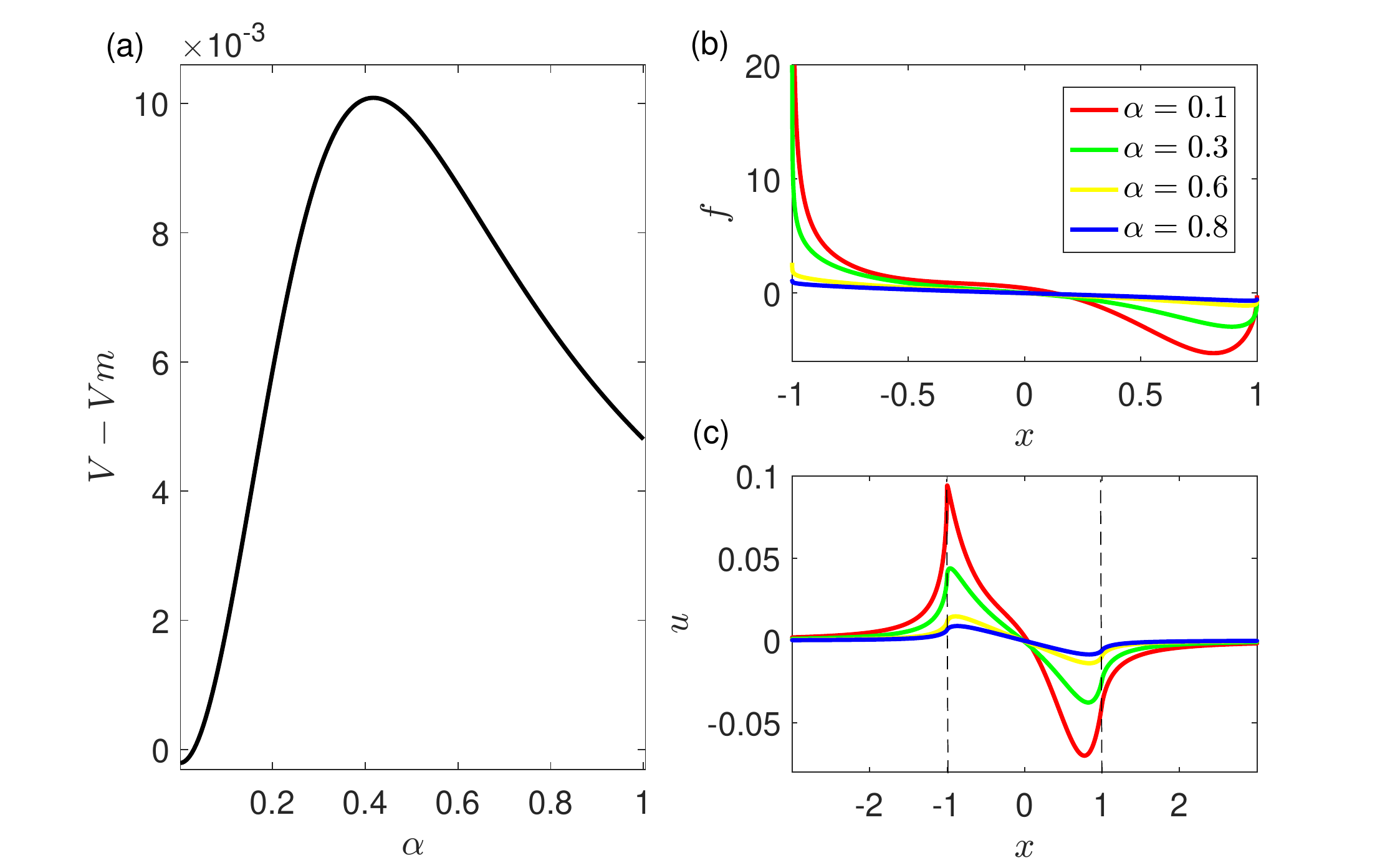}
\caption{\label{fig:V_alpha} \textcolor{red}{ Dependence of the crawling velocity on the slip coefficient with the substrate. (a) shows the biphasic regime as a function of $\alpha$. Traction forces (b) and substrate displacements (c) are given for some specific values of $\alpha$.  Parameters are $\epsilon=0.2$, $\gamma=0.02$, $\Delta V=1$ and $V_m=1.5$.}}
\end{figure}}
\textcolor{red}{The crawling velocity cannot be directly related to the magnitude of the traction force or the mechanical work performed on the substrate as the odd component of $f$ does not contribute to motility. However, we do observe that the  $V(\gamma)$ curve is associated with a biphasic behavior of the work of the traction forces $W=\int_{-1}^{1}f(x)u(x)\mathrm dx$ while their magnitude $I=\int_{-1}^{1}|f(x)|\mathrm dx$ keeps decreasing as $\gamma$ increases. This is consistent with  Fig.~\ref{fig:V_gamma} where we observe that the magnitude of traction forces decreases while the one of the substrate displacement increases when $\gamma$ varies between $0.01$ and $0.02$, which corresponds to the biphasic region for $V$. We quantity this effect on Fig.~\ref{fig:traction_intensity_work}, where we show the variation of $I$ and $W$ as a function of $\gamma$. 
\begin{figure}[!ht]
\centering
\includegraphics[width=0.5\textwidth]{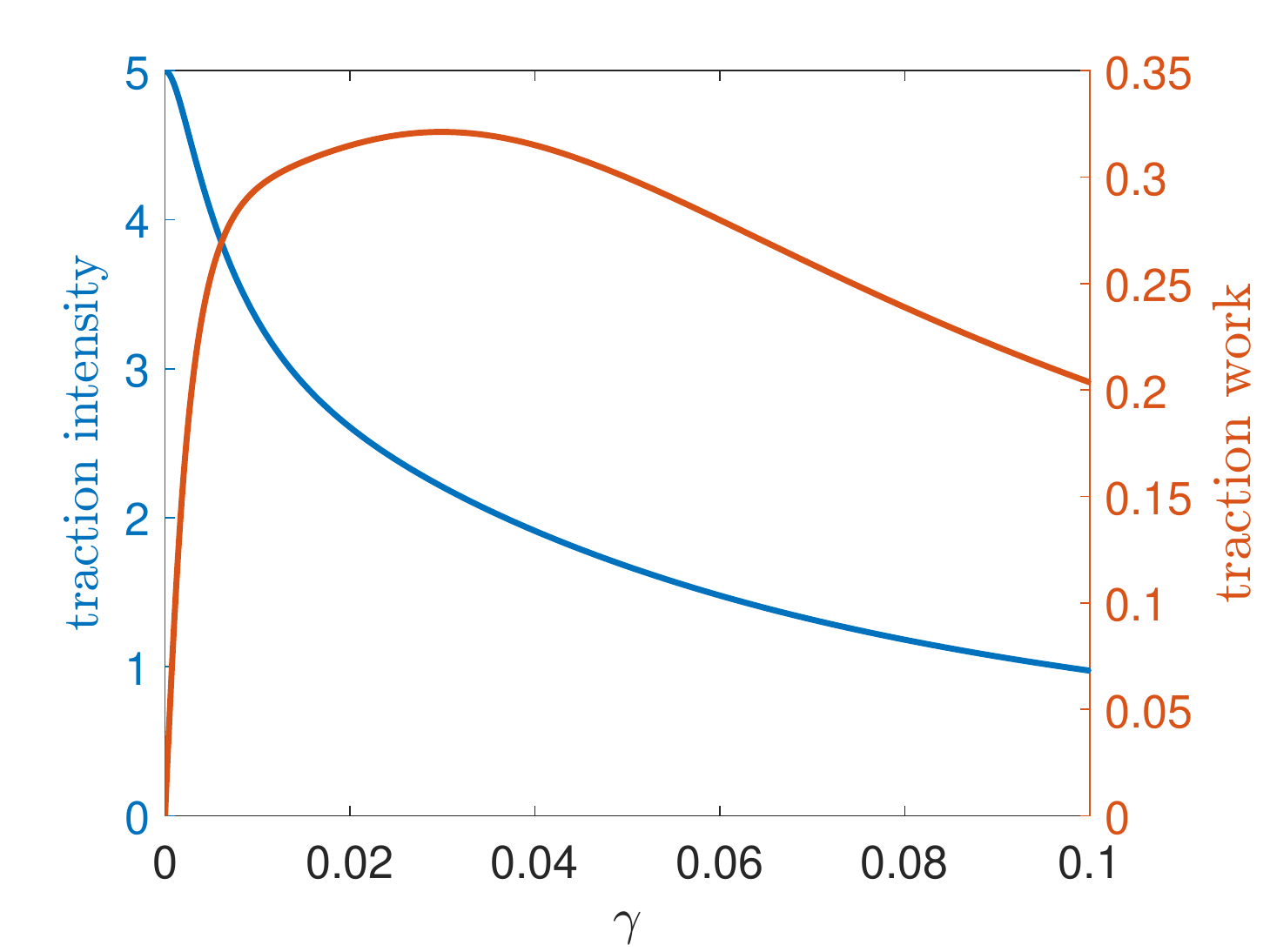}
\caption{\label{fig:traction_intensity_work} \textcolor{red}{ Evolution of the traction force magnitude $I=\int_{-1}^{1}|f(x)|\mathrm dx$ and traction force work $W=\int_{-1}^{1}f(x)u(x)\mathrm dx$ as a function of $\gamma$ at a fixed $\alpha$.    Parameters are $\epsilon=0.2$, $\alpha=0.1$, $\Delta V=1$ and $V_m=1.5$.}}
\end{figure}
A plausible explanation of the present biphasic  regime of the velocity in $\gamma$ is therefore that while the traction forces decrease with the substrate stiffness, their work is maximal at a certain value of $\gamma$ leading to an optimal propulsion at least at a local level close to this value.
}

\textcolor{red}{But this explanation does not hold for the global biphasic behavior in $\alpha$ shown in  Fig.~\ref{fig:V_alpha} where both the magnitude of the traction forces and their mechanical work on the substrate keep decreasing as $\alpha$ increases. In this case, the importance of the substrate stiffness is  to regularize the behavior of the velocity when $\alpha$ is small. Indeed, in the limit where the substrate is very stiff, the velocity expansion blows up when $\alpha\ll 1$  as we show in Fig.~\ref{fig:V_1_alpha}, indicating that the expansion is no longer valid in this regime. A finite $\gamma$ is sufficient to restore a finite value of the speed when $\alpha\rightarrow 0$ (see Fig.~\ref{fig:V_alpha}) and regularizes the problem. Physically, $\alpha\ll 1$ corresponds to a large friction coefficient where outside of sharp boundary layers, $v_x=v_x^s$ such that the cytoskeleton velocity is transmitted to the substrate. It is therefore expected that the substrate deformability is playing an important role in fixing the velocity in this regime. The other limit where $\alpha\gg 1$ corresponds to a vanishing friction coefficient between the cell and the substrate and the two problems therefore uncouple. We can thus expect that $V=V_m$ in this regime as the substrate deformation stops contributing to the crawling motion. In between these two regimes, Fig.~\ref{fig:V_alpha} shows that there exists a friction coefficient which maximizes the crawling velocity as traction forces are transmitted to the substrate contributing to the propulsion while the coupling with  the substrate is minimally braking the motion.}

\textcolor{red}{Contrary to the biphasic regime with respect to the substrate stiffness, the biphasic regime with respect to the substrate friction coefficient is not due to the precise constitutive behavior of the substrate. To demonstrate this, we consider in \ref{sec:kernel_exp} an exponential kernel replacing $\Phi(x)$ by $\Phi_{\text{exp}}(x)\sim\exp(-|x|)$. Such non-singular kernel mimics the non-local feedback of the substrate on the cell motion but is of a shorter range compared to $\Phi(x)$ as it implies that the characteristic lengthscale of the displacement decay in response to a point force is $\sim \epsilon$ while it is $\sim 1$ for the elastic kernel. In this case, while we recover a generic biphasic regime for the $V(\alpha)$ curve, the $V(\gamma)$ curve is monotonically increasing. }

\section{Experimental platform}\label{sec:platform}

\begin{figure}[ht]
    \centering
    \includegraphics[scale = 0.3]{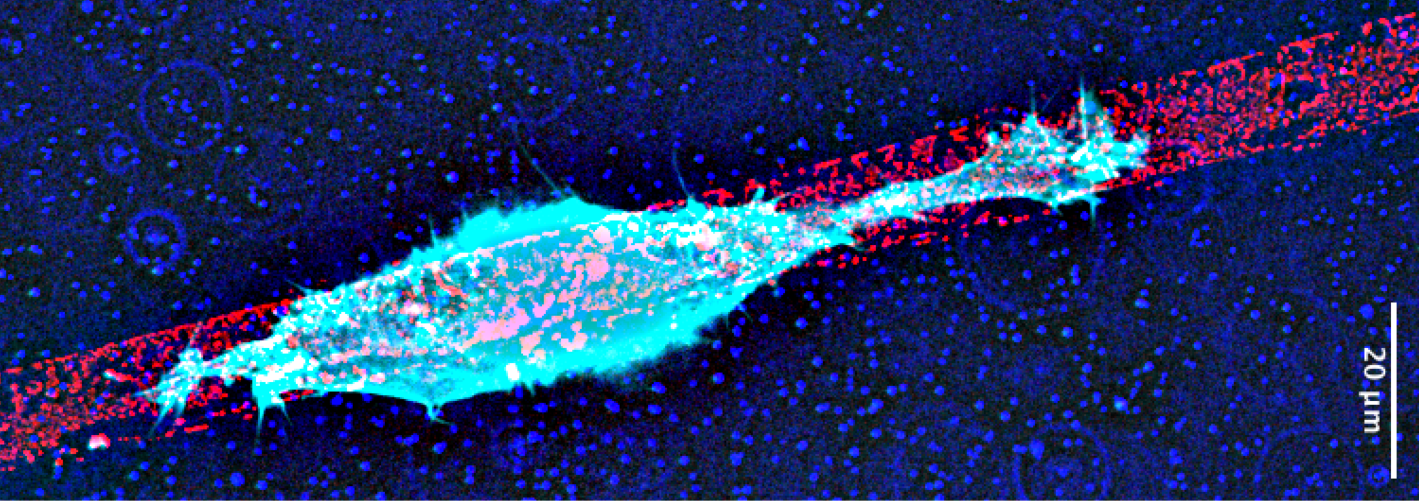}
    \caption{Fluorescent cell (actin in cyan) migrating on a fibronectin track (red) $w \sim 13\,\mu$m on a gel with embedded beads (dark blue).}
    \label{fig:F}
\end{figure}

We then aimed at constructing an in-house experimental platform to investigate in details the influence of the substrate rigidity on the cell motion and check the range of validity of our simple one-dimensional theoretical model. To do so, it is necessary to produce a substrate with a suitable pattern and appropriate stiffness. To answer this request, patterning methods and cell-compatible hydrogels have been developed that are able to create protein micropatterns on substrates with varied stiffness \citep{grevesse2014preparation,grevesse2013simple}. Hydroxy-PAAm(polyacrylamide) are good candidates and can be functionnalized with fibronectin. After cell seeding on such hydrogels, it is possible to acquire microscopic images of the cells and track fluorescent beads embedded within the substrate in order to study the relationship between cell migration and substrate deformation. Added to this, we used T24 epithelial bladder cancer cells which have been transfected with actin--GFP \citep{peschetola2013time}, so that the actin velocity can be measured. We can therefore image at the same time the substrate beads and the actin meshwork as shown in Fig.\ref{fig:F} to estimate the substrate displacement and the actin velocity.  The necessary steps to follow are detailed in \ref{sec:exp}.
\textcolor{red}{In a real experimental context, the substrate is not semi-infinite but has a finite thickness. We justify the use of such approximation in \ref{sec:semi_inf_approx}}

\begin{figure}[htb]
    \centering
    \includegraphics[scale = 0.45]{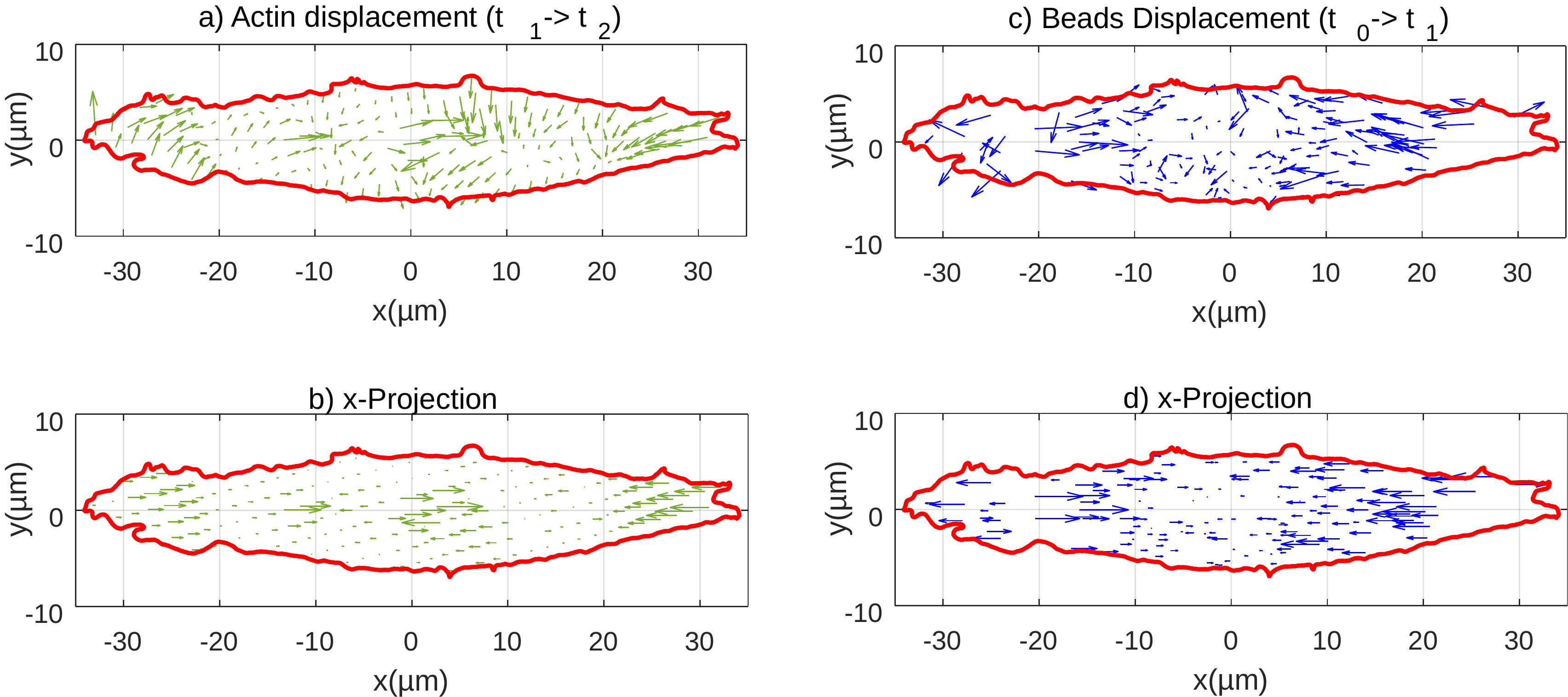}
    \caption{ (a) Actin displacement between two successive acquired time frames ($t_2$ and $t_1$ are separated by 5 s) and (b) its projection along the track direction. (c) Beads displacements between a given time frame ($t_1$) and a reference configuration where the cell is absent (after detachment) ($t_0$). The projection along the track axis is shown on (d). \textcolor{red}{Our assumption \eqref{eq:traction2d} that the traction forces only depends on the $x$-coordinate is not well satisfied in our experiments and {we will need} to produce thinner tracks in the future for a better applicability of our one-dimensional model.} }
    \label{fig:Ro_Pr}
\end{figure}

In order validate our 1D-model, \textcolor{red}{since experimentally attainable tracks that are not thin enough for our one-dimensional assumption \eqref{eq:traction2d} to be fully adequate,} we project the actin velocity and the substrate displacement on the major axis of the cell, where the origin of the axis corresponds to the cell center (see Fig.~\ref{fig:Ro_Pr}).
For a reason that is not yet clear to us, our cells were poorly motile and we therefore restricted our analysis to an acquisition sequence of 15 time frames with a 5 s time lag between them where the cell remains almost static. In this situation, we measured an incremental displacement of the substrate of the order of the measurement uncertainties, corresponding to the resolution of the measurement where $\partial_t u_x^s\ll v_x$.  Therefore \eqref{eq:disp1D_SIDE} becomes a simple integral equation relating $u_x^s$ and $v_x$:
\begin{equation}\label{e:us_static}
  u^s_x(x) = \frac{\xi L}{2}\int_{-1}^{1}\phi\left(\frac{x'-x}{\epsilon}\right)v_x(x')\mathrm dx',
\end{equation}
Due to the resolution of the measurements, a linear approximation of the actin retrograde flow velocity is appropriate, as pushing the approximation to higher order would essentially capture noise. Thus we write $v_x(x)\simeq v_1 x$, where the coefficient $v_1$ can be physically interpreted as the slope of \eqref{e:v_0} at the origin. In order to reduce the effect of fluctuations, a linear regression is performed at every time frame and then averaged over the total number of frames:
$$v_x(x) = \frac{1}{N}\sum_{n=1}^N v_1^n x,$$
where $N$ is the total number of time frames in a sequence ($N=15$) and $v_1^n x$ is the linear regression of the actin velocity for frame $n$.
\begin{figure}
     \begin{subfigure}[b]{0.32\textwidth}
         \includegraphics[height=5cm]{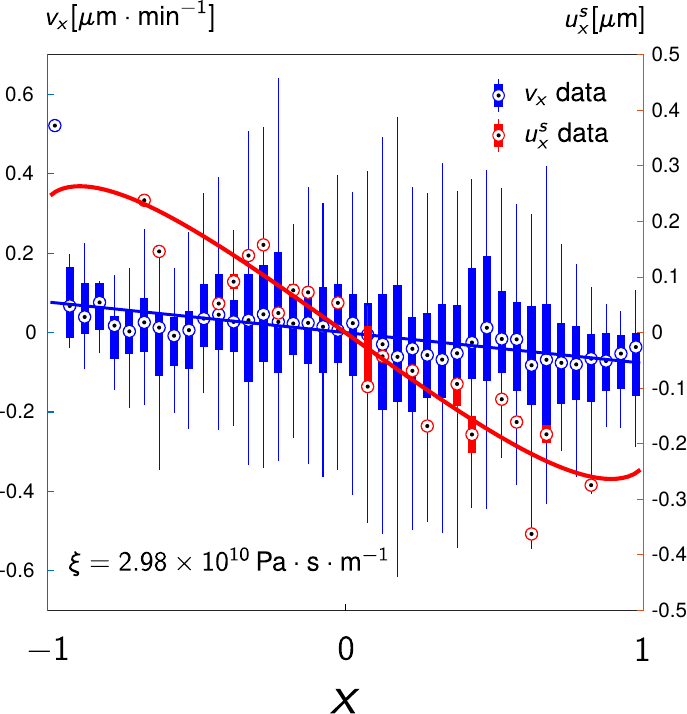}
         \caption{$E_s = 5$ kPa}
         \label{fig:y equals x}
     \end{subfigure}
      \hfill
     \begin{subfigure}[b]{0.32\textwidth}
         \includegraphics[height=5cm]{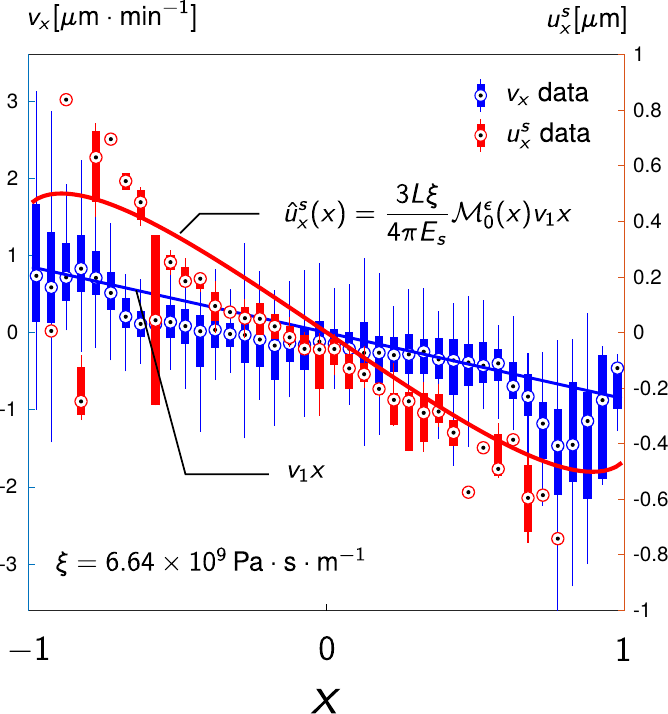}
         \caption{$E_s = 8$ kPa}
         \label{fig:three sin x}
     \end{subfigure}
      \hfill
     \begin{subfigure}[b]{0.32\textwidth}
         \includegraphics[height=5cm]{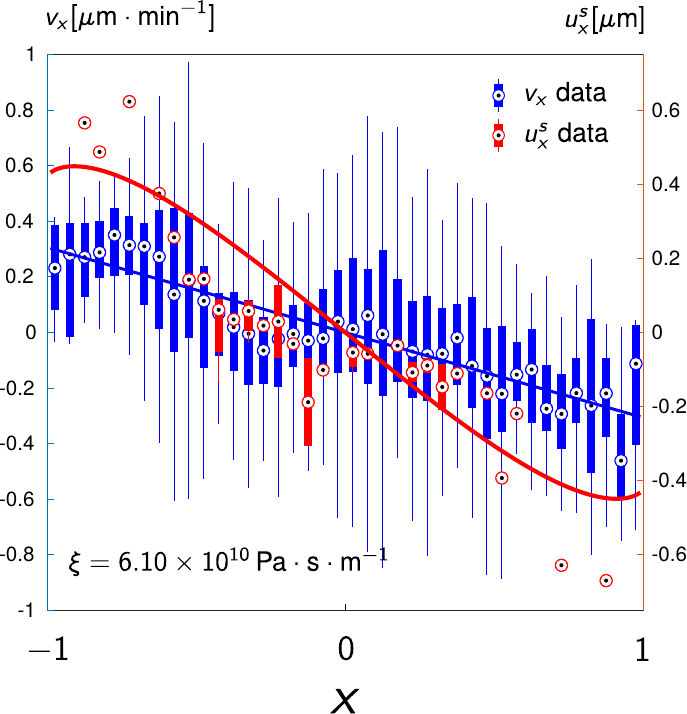}
         \caption{$E_s = 28$ kPa}
         \label{fig:five over x}
     \end{subfigure}
        \caption{Estimation of the friction coefficient $\xi$ from the experimental data for three different substrate rigidities (5 kPa, 8 kPa, 28 kPa). Over one sequence (15 frames with a 5s timestep), the measured actin velocity is represented using 40 blue boxplots and the measured substrate displacement under the cell using 40 red boxplot. Each boxplot is computed considering all the data points in 1/40 of the cell. The blue line represents the linear regression of the measured actin velocity over the sequence. The red line is the predicted substrate displacement obtained using our model  \eqref{us_predicted_ss}, where the value of $\xi$ is extracted by performing a least square minimization between the predicted (red line) and measured (red boxplots) substrate displacements.}
        \label{fig:xi_estimation}
\end{figure}
From \eqref{e:us_static}, the substrate displacement for $x\in[-1,1]$ for the linearized expression of $v_x$ reads,
\begin{equation}\label{us_predicted_ss}
    u^s_x(x) = \frac{3 L\xi}{4\pi E_s}v_1 x\mathcal{M}_0^\epsilon(x).
\end{equation}
where,
\begin{align*}
\mathcal{M}^\epsilon_{0}(x)=& \epsilon\, \text{arcsinh}\left(\frac{1-x}{\epsilon }\right)+\epsilon\, \text{arcsinh}\left(\frac{1+x}{\epsilon }\right)\\
+&(1-x)\, \mathrm{log}\left( \frac{\epsilon+\sqrt{(1-x)^2+\epsilon^2}}{1-x}\right)+(1+x)\, \mathrm{log}\left( \frac{\epsilon+\sqrt{(1+x)^2+\epsilon^2}}{1+x}\right).
\end{align*}

We then estimate $\xi$  by minimizing the  distance between the predicted substrate displacement \eqref{us_predicted_ss} for the measured actin retrograde flow velocity and the substrate displacement from the experimental data. The results for three different substrate rigidities ($E_s$ = 5 kPa, 8 kPa and 28 kPa) are shown in Fig.~\ref{fig:xi_estimation}. In these plots the actin velocity and substrate displacement experimental data are represented using boxplots. We observe strong fluctuations of the actin velocity for the 5 kPa and 28 kPa rigidities, while the data points are much more concentrated in the 8 kPa case. This is due to the fact, that in the cases of 5 kPa and 28 kPa, the cells where not elongated enough (not 1D), thus at the same abscissae, we could measure two opposite velocities in some cases. The actin retrograde flow was also difficult to track because of the diffuse fluorescence. 

While the friction coefficient seems to exhibit a slight dependency to the substrate rigidity, ranging from $6.64\times10^{9}\, \text{Pa.s.m}^{-1}$ to $6.10\times10^{10}\,  \text{Pa.s.m}^{-1}$, more data are needed to extract a clear tendency, as only one cell sequence was retained for each substrate rigidity. Because $\partial_t u_x^s\ll v_x$, the viscous friction law \eqref{eq:viscous_friction} reduces to $T_x(x,t) = \xi v(x,t)$, which corresponds to the friction law on a rigid substrate. This gives us a comparison point to evaluate the relevance of our model. Indeed, while obviously the substrate displacement between soft and rigid substrates cannot be compared (being zero in the latter), the friction law is identical in this specific case. Therefore to validate our model the friction coefficient for a compliant substrate has to be of the same order as the one for a rigid substrate, which we note $\xi_r$. Extended work has already been done to characterize cell crawling on rigid substrates by mean of a viscous friction law \citep{Juelicher2007, Rubinstein2009} and $\xi_r$ has been evaluated to be of the order of $10^{9}-10^{10}\, \text{Pa.s.m}^{-1}$. With our model we successfully recover a friction coefficient of the same order of magnitude as the one obtained on a rigid substrate. 

With this validated experimental setting, we plan to investigate in details situations where the cell is strongly motile in the future.

\section{Conclusion}
 
Starting from a linear elastic semi-infinite substrate and a viscous friction law linearly relating the cell traction forces and the relative velocity between the actin and the substrate, we modeled the mechanical interaction between a crawling cell and a compliant substrate for cells confined to move on thin micropatterned fibronectin tracks. This model of the cell contact was then coupled with one of the simplest model of cell propulsion based on actin turnover. In such a model, the polymerization at the cell front and its depolymerization at the cell back leads to two motility mechanisms. The addition and removal of actin monomers induces a certain treadmilling velocity that is independent of external conditions but also a retrograde flow of actin from the cell front to its back exists and is dependent on the mechanical coupling with the substrate. Because of such non-local coupling, we find that the dependence of the cell velocity on either the substrate stiffness at a given friction coefficient or the friction coefficient at a given substrate stiffness \textcolor{red}{contain a biphasic regime for a realistic choice of the model parameters}. This offers an interesting paradigm to complement other theories directly invoking a local non-linear dependence of the friction force on the actin flow to explain this global non-monotonicity. Experimentally, through the simultaneous monitoring of the substrate displacement and the actin retrograde flow velocity for static cells, we were able to extract the effective friction coefficient that enters in our model for different substrate rigidities by performing model-based data-fitting. We observed a small variation of the friction coefficient with the substrate rigidity, but the values globally lie in the same magnitude range. We find values of the friction coefficient of the same order as the ones  previously reported for  cells crawling on a rigid substrate suggesting that it is legitimate to consider this parameter as fixed regardless of the substrate rigidity. In the case of the 5 kPa and 28 kPa subtrate rigidities, the cells were not elongated enough to consider them as one dimensional and the actin acquisition resolution was poor. To address these issues it would be necessary to be able to pattern thinner fibronectin tracks.

We expect our experimental method to be of greater interest for cell types and substrate rigidities where the actin retrograde flow and the rate of deformation of the substrate are comparable and the cells are more motile than in the present work. Another perspective is to couple the present model of non-local cell-to-substrate contact to the paradigmatic case of contraction-driven cell motility \citep{Recho2015} where we expect complex intermittent gaits to appear from the coupling of the Keller-Segel instability to the non-locality induced by the substrate elasticity. 

\section{Acknowledgements}
We thank Alain Duperray for providing the transfected cells, and for related discussions. We acknowledge support from a CNRS MOMENTUM grant and the ANR (grant No. 12-BS09-020-01, TRANSMIG). C.V., J.E. and P.R. are members of the LabeX Tec 21 (Investissements d’Avenir: grant agreement No. ANR-11-LABX-0030).

\appendix 

\section{\textcolor{red}{Interaction kernel with the substrate with exponential decay}} \label{sec:kernel_exp}

\textcolor{red}{We formulate the new problem where we replace the expression of $\Phi$ in \eqref{e:v_turnover} by $\Phi_{\text{exp}}(x)=A\text{e}^{-|x|}$ where we set the constant
$$A(\epsilon)=\frac{e^{\frac{1}{\epsilon }} \left(\epsilon  \log \left(\frac{\sqrt{\epsilon ^2+1}+1}{\epsilon }\right)+\sinh ^{-1}(\epsilon )\right)}{\left(e^{\frac{1}{\epsilon }}-1\right) \epsilon },$$
such that $\int_{-1}^{1}\Phi(x/\epsilon)\mathrm dx=\int_{-1}^{1}\Phi_{\text{exp}}(x/\epsilon)\mathrm dx$. With such exponential kernel, system \eqref{e:v_turnover} can be written in differential form:
\begin{equation}\label{e:exp_kernel_syst}
\left\lbrace \begin{array}{c}
-\alpha^2\partial_{xx}v_x+v_x=-V\partial_xu\\
-\epsilon^2\partial_{xx}u+u=2\epsilon A\gamma\partial_{xx}v_x
\end{array}\right. \text{ with boundary conditions }
\left\lbrace\begin{array}{c}
\partial_xv_x\vert_{-1}=\partial_xv_x\vert_{1}\text{ and }v_x\vert_{1}-v_x\vert_{-1}=-\Delta V\\
\epsilon\partial_xu\vert_{-1}-u\vert_{-1}=0 \text{ and }\epsilon\partial_xu\vert_{1}+u\vert_{-1}=0
\end{array}\right. 
\end{equation} 
and the unknown velocity is still fixed by the condition
$$V=V_m+\frac{v_x\vert_{-1}+v_x\vert_{1}}{2}.$$ 
System \eqref{e:exp_kernel_syst} is a fourth order boundary value problem with a free parameter $V$ that is set by the previous condition. Solving this problem using a continuation method, we construct the $V(\gamma)$ and $V(\alpha)$ curves that we superimpose with their analogue for the elastic kernel $\Phi$ on Fig.~\ref{fig:exp_elastic_comp}.
\begin{figure}[!ht]
\centering
\includegraphics[width=1\textwidth]{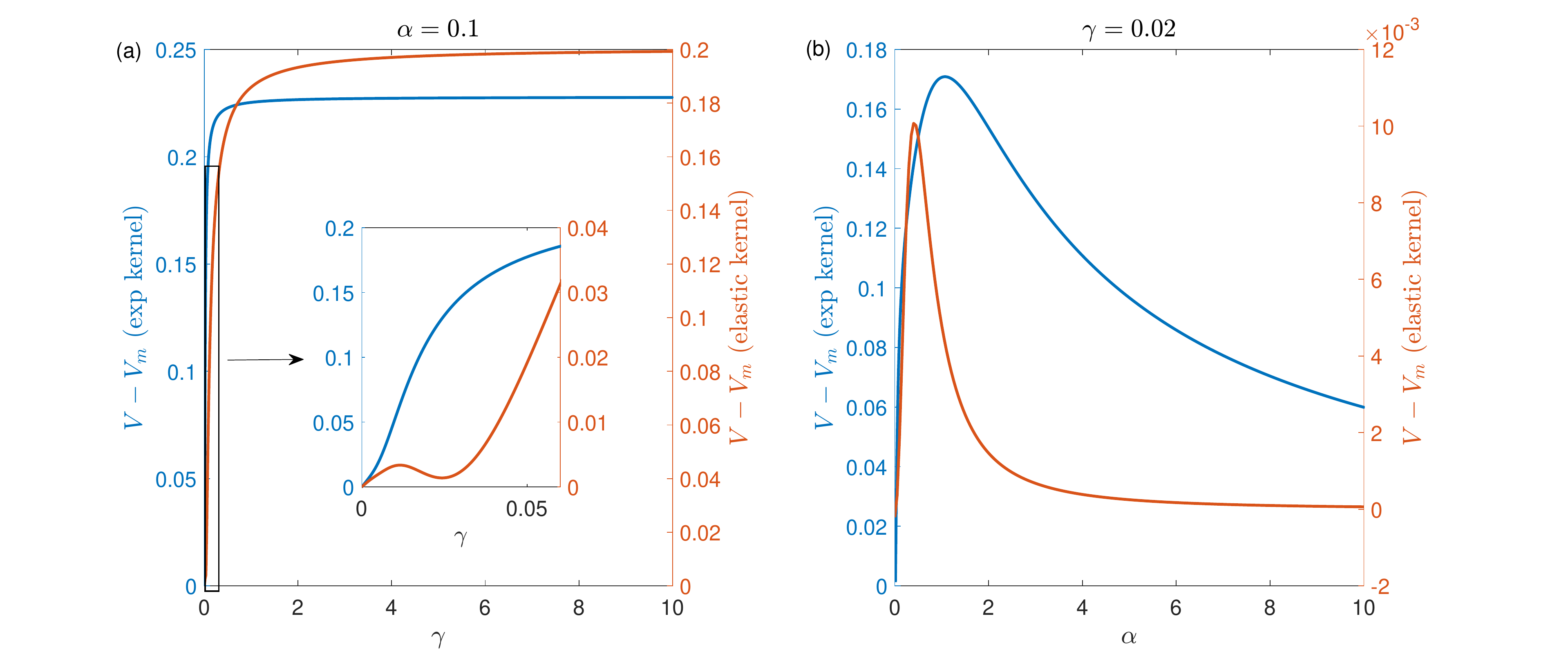}
\caption{\label{fig:exp_elastic_comp} \textcolor{red}{ Dependence of the crawling velocity on (a) the substrate softness $\gamma$  and (b) the substrate slip coefficient $\alpha$ for both the exponential and elastic interaction kernel with the substrate.  Parameters are $\epsilon=0.2$,  $\Delta V=1$ and $V_m=1.5$.}}
\end{figure}
We observe that while the global biphasic structure in $\alpha$ is still present, the local one in $\gamma$ disappears.}

\section{Experimental and post processing methods}
\label{sec:exp}

$\blacktriangleright$ \textsc{\textbf{Synthesis of hydroxy-PAAm hydrogels}}

\begin{figure}[htb]
    \centering
    \includegraphics[scale = 0.7]{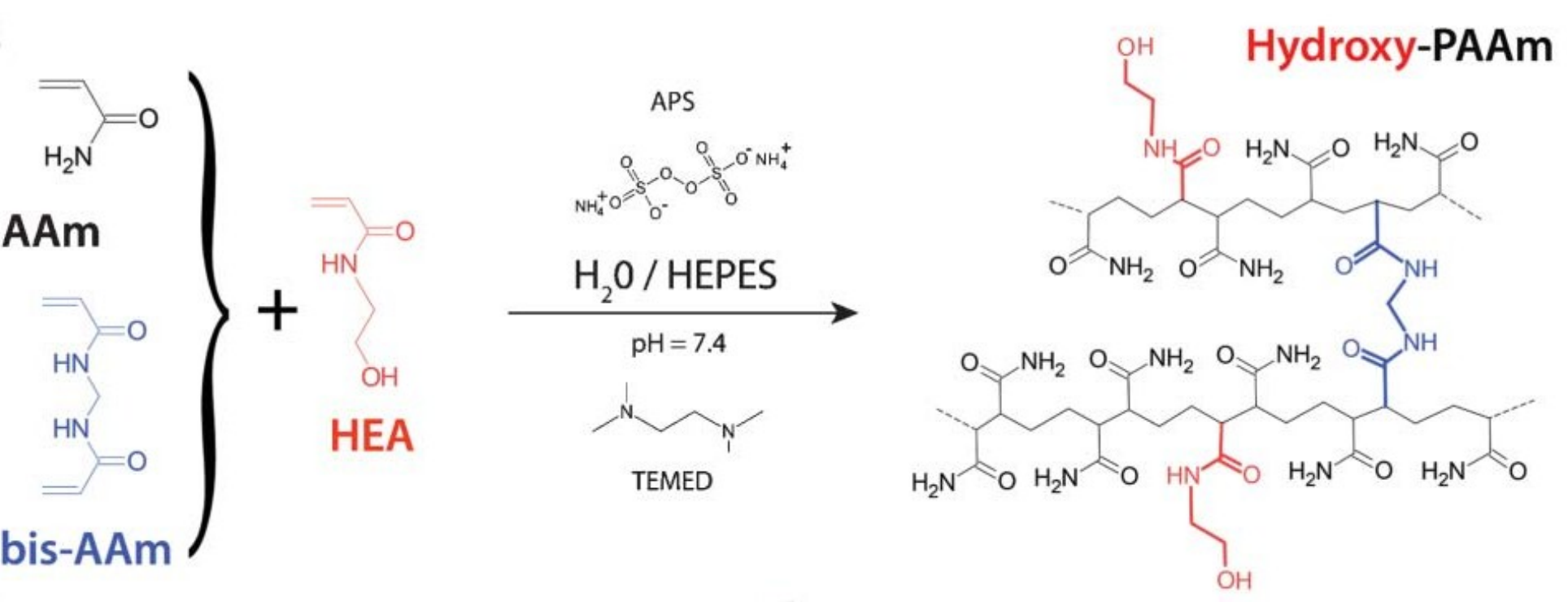}
    \caption{AcrylAmide (AAm, in black), N,N$^{'}$ methylene--bis--AcrylAmide (bis-AAm, in blue) and N-Hydroxy--EthylAcrylamide (HEA, in red) were mixed together to form hydroxy-PAAm \cite{grevesse2013simple}.}
    \label{fig:chemi}
\end{figure}

In order to prepare hydroxy-PAAm gels to be coated with fibronectin for cell adherence, it is necessary to use acrylamide (AAm) mixed with N-hydroxyethylacrylamide (HEA) monomers. The latter contains  HO groups that can form a hydrophilic network (see Fig. \ref{fig:chemi}). Then using bis--acrylamide as a crosslinker and TEMED/APS for catalysis, rapid polymerization of acrylamid monomers is obtained \cite{grevesse2014preparation,abidine2018}.

Here, we used the protocol explained in \cite{abidine2018}. We prepared solutions by mixing acrylamide (30\% weight per weight [w/w]), N-hydroxyethyl-acrylamide (5.85\% w/w), and N,N-methylene-bisacrylamide (2\% w/w) in different amounts (Sigma-Aldrich, St. Louis, MO). Three concentrations of bis--acrylamide were used (0.1, 0.3, and 0.6 \%), with the acrylamide (3.2\%) and N-hydroxyethyl-acrylamide (1.25\%) contents remaining fixed in the final HEPES solution (50 mM). Gels were 70 $\mu$m thick with an area 1.5 cm $\times$ 1.5 cm, and were prepared on a glass slide (pre--treated with 3--Aminopropyl--triethoxysilane, APTMS) glued at the bottom of a Petri dish.

\begin{figure}[htb]
    \centering
    \includegraphics[scale = 1.2]{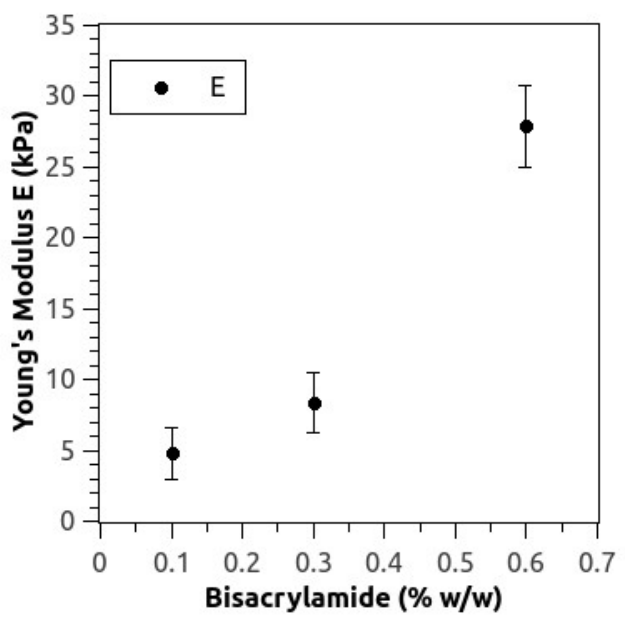}
    \caption{The stiffness of hydroxy-PAAm hydrogels is increasing with the amount of bis-AAm cross-linker \cite{Abidine2015Pro}.}
    \label{fig:elastic}
\end{figure}

The gel rigidity should be chosen as a compromise between the stiffness in physiological conditions and the sufficiently large displacements to be measured in soft gels (on the order of a few $\mu$m). Values of elastic modulus ($E$) between 5kPa and 30 kPa were found to be adequate \cite{peschetola2013time}. The elastic Young moduli of the three hydrogels were measured using an Atomic Force Microscope (JPK AFM, NanoWizard II, Berlin) in contact mode, equipped with MLCT cantilevers (pyramidal tips, stiffness 0.01 N/m, Bruker). Five different locations were selected and 5x5 elasticity maps were performed at each location. The values were obtained using the classical relationship 
$F =\frac{3}{4}\frac{E}{1-\nu^2}\,\text{tan}\theta\,\delta^2$, where $F$ is the applied force, $\nu \sim 0.5$ is the Poisson ratio, $\theta$ the half--pyramid angle, and $\delta$ is the indentation. The resulting elastic moduli were found to be 5 $\pm$ 1 kPa, 8 $\pm$ 1.5 kPa and 28 $\pm$ 3 kPa. This is shown in Fig. \ref{fig:elastic} with hydrogel elastic moduli increasing with cross-linker concentration as previously measured \cite{Abidine2015Pro}.

\vspace{0.3cm}

$\blacktriangleright$ \textsc{\textbf{Fibronectin micropatterning of hydroxy-PAAm gels}}

\begin{figure}[ht]
    \centering
    \includegraphics[scale = 0.7]{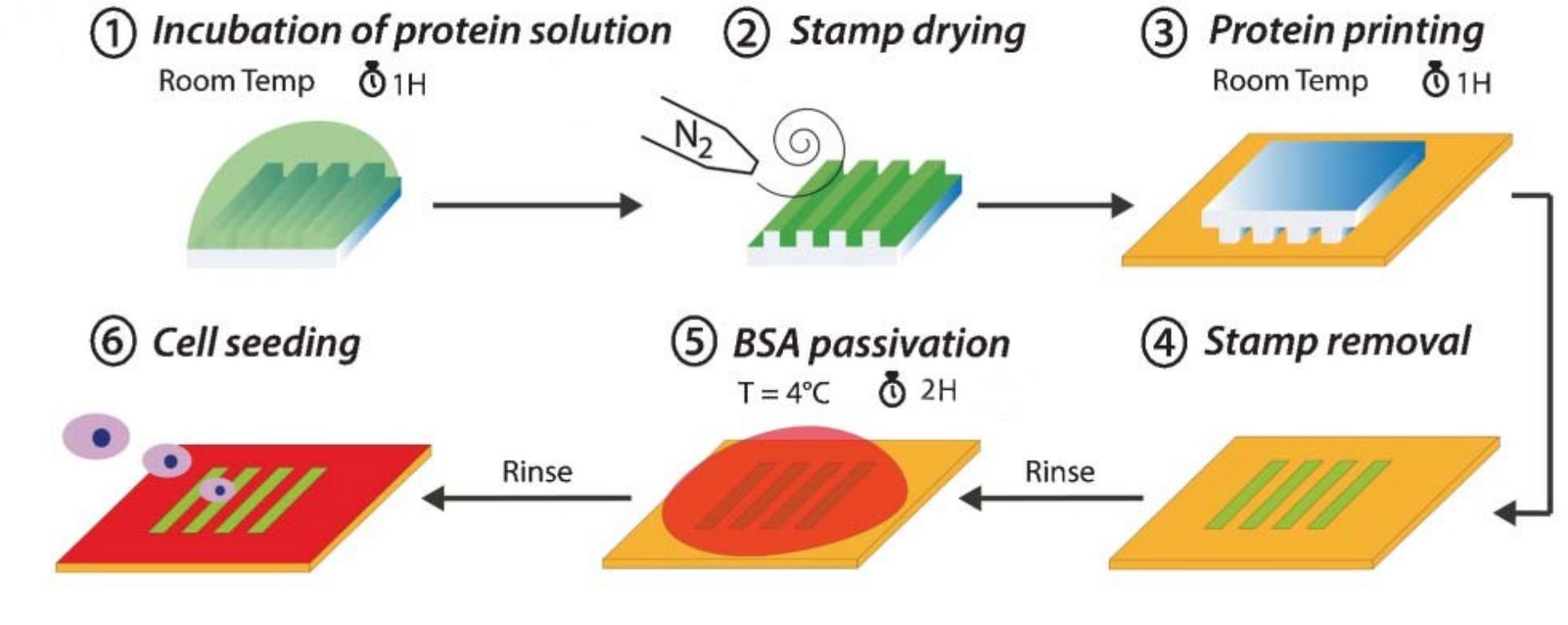}
    \caption{Process of fibronectin micropatterning on hydrogel \cite{grevesse2013simple}.}
    \label{fig:stamp}
\end{figure}

Fibronectin (FN) mediates a wide variety of cellular interactions with the ECM and plays an essential role in cell adhesion, migration, growth, and differentiation \cite{pankov2002fibronectin}. Incubation of fibronectin straight microtracks with a very narrow width (less than $15\, \mu$m) and far from each other (more than $40\,\mu$m) on micro-fabricated hydroxy-PAAm hydrogels can be used to simulate 1D migration of cells on soft substrates. Here we used a fibronectin micropatterning process by designing the required PDMS\footnote{Polydimethylsiloxane} stamp \cite{grevesse2013simple} as shown in Fig. \ref{fig:stamp}.


\begin{figure}[ht]
    \centering
    \includegraphics[scale = 0.6]{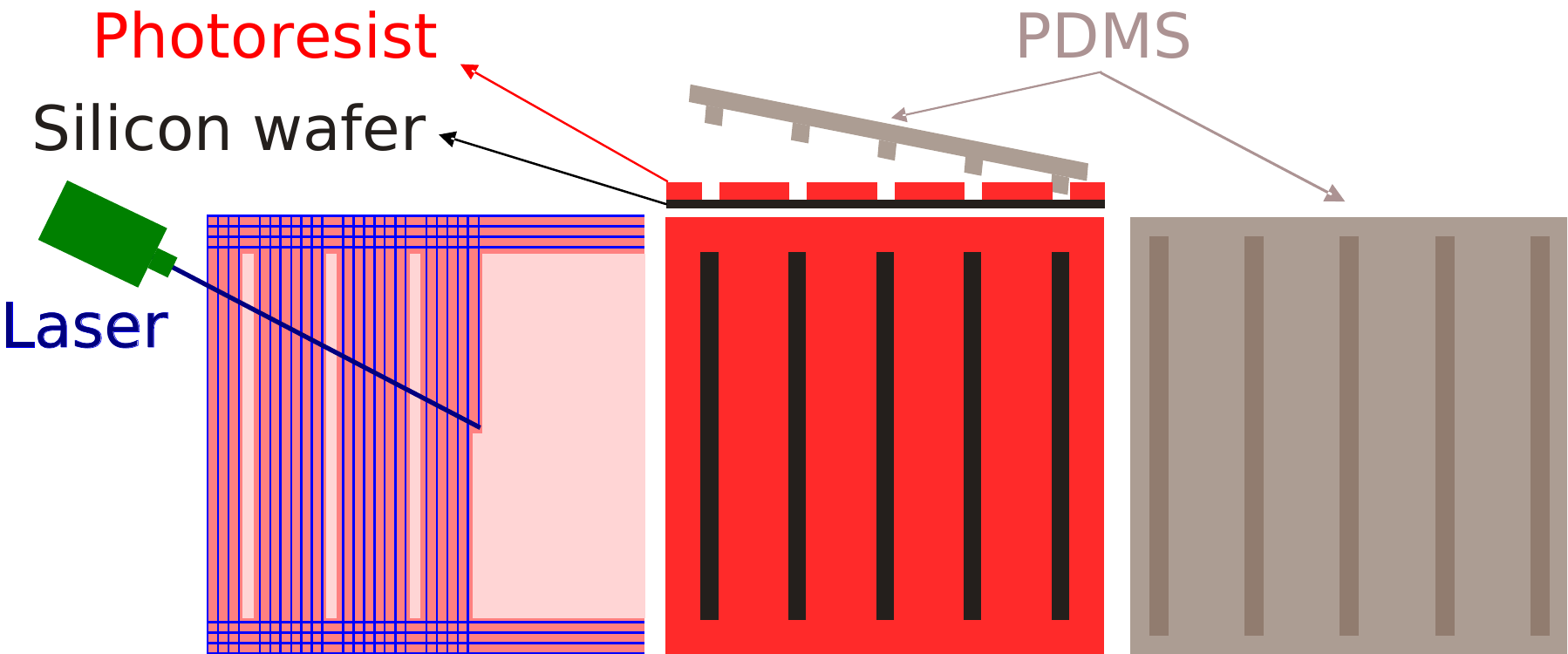}
    \caption{Process of the PDMS stamp fabrication}
    \label{fig:laser}
\end{figure}

For the preparation of patterns on PDMS, we first made a silicon master using optical lithography with a negative photoresist. Optical lithography is a photographic method by which a light-sensitive polymer, named a photoresist, is exposed and developed to form 3D features on the silicon substrate. A negative photoresist is one whose UV exposed parts become cross-linked and other parts remain soluble and can be washed away during development. One of the commonly used epoxy--based negative photoresists is SU-8 GM1070, a chemically amplified resist system with excellent sensitivity. The final micropattern presents regions with photoresist and other parts are uncovered. This coated micropattern is needed to shape the PDMS. The common steps for a regular photolithography process are as follows: substrate preparation, photoresist spin coating, prebaking, exposure, post-exposure baking, development, and post-baking \cite{mack2016semiconductor}. In order to have the silicon master with about $15\,\mu$m in height, we followed the SU8-Photoepoxy GM 1070 datasheet.
The essential principle behind this photoresist operation is the change in the solubility of the photoresist in the developer, upon exposure to light. Here we used UV laser exposure with $10 \,\mu$m laser beamwidth in order to change the solubility of the photoresist and as a result we obtain designed photoresist micro-patterns on top of the silicon wafer (Fig. \ref{fig:laser}). Then PDMS is poured onto the substrate, baked at 60$^{\circ}$C for 2h, and is finally peeled off.

According to Fig. \ref{fig:laser}, by giving a designed pattern to the laser machine, it is possible to make our micro--pattern directly on the photoresist. There are two essential parameters, the percentage of exposure energy and the time of exposure. The time of exposure comes from the velocity of the moving laser beam. By testing different exposure energy percentages and velocities, we found that 100 $\%$ and 25 mm/s give the best results for 15 $\mu$m height of photoresist layer on the substrate.

\begin{figure}[ht]
    \centering
    \includegraphics[scale = 0.5]{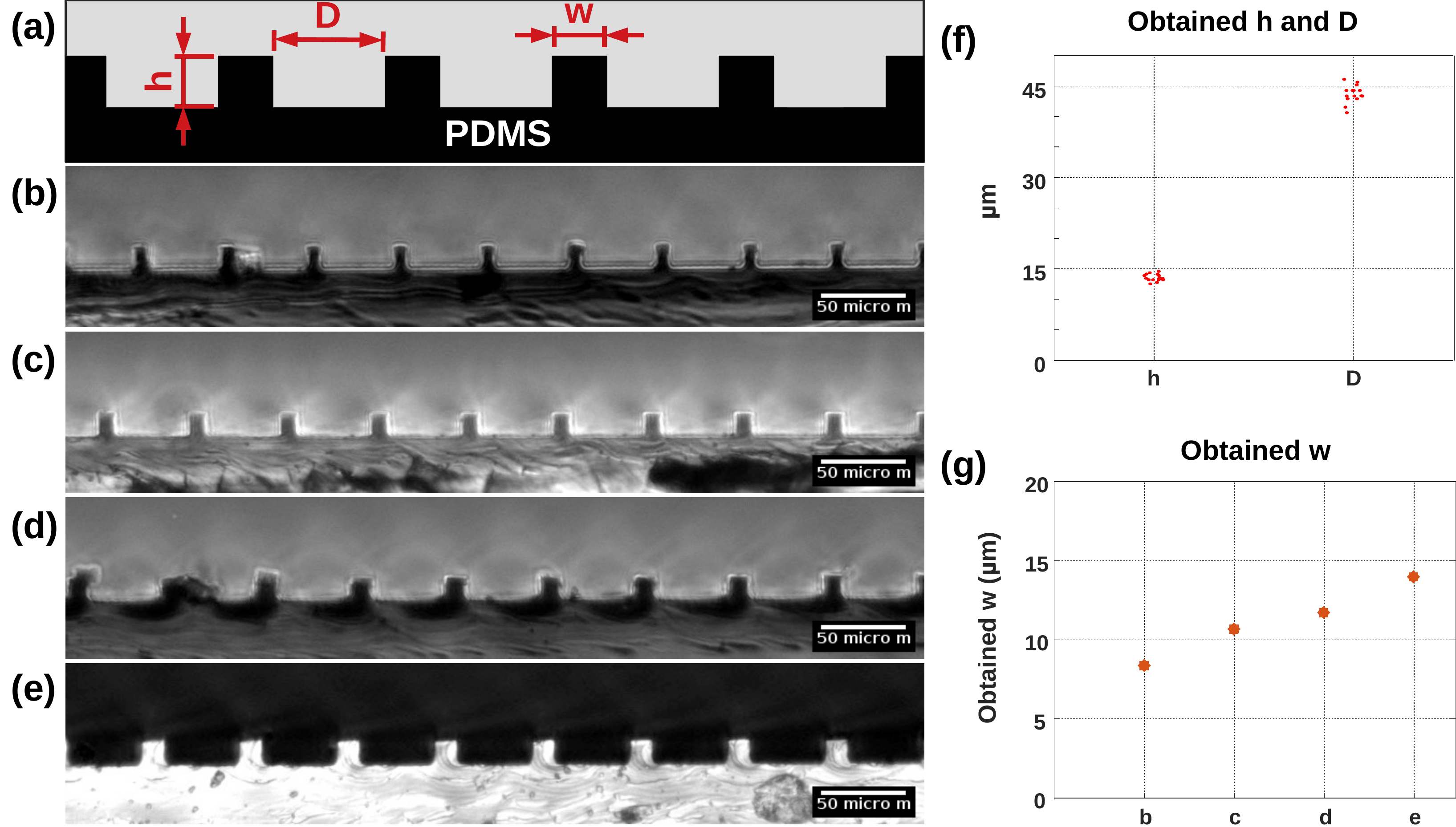}
    \caption{a) Schematic of a cross-section of PDMS stamp with $w$ (width of lines), $h$ (height of lines), and $D$ (distance between two lines) b, c, d, e) Images of PDMS cross-sections with different $w$ f) Distribution of measured values $h\simeq15 \,\mu$m and $D \simeq 45 \,\mu$m. g) Values of $w$: 8, 11, 12, and 14 $\mu$m for images b, c, d, and e respectively.}
    \label{fig:PDMS}
\end{figure}

In order to check the final shape of designed PDMS stamps with different patterns (Fig. \ref{fig:PDMS}a), we cut very thin cross-section layers of PDMS and imaged them under an optical microscope (Fig. \ref{fig:PDMS}b--e). Then we measured the width ($w$) and height ($h$) of the lines, as well as the distance between two lines ($D$) using Fiji\textsuperscript{TM} software. According to Fig. \ref{fig:PDMS}f, the results show that $h\simeq15 \,\mu$m and $D \simeq 45 \,\mu$m are almost similar for all PDMS with different $w$. Also, Fig. \ref{fig:PDMS}g shows that the smallest $w$ that we could obtain was $w \simeq 8 \,\mu$m. Although for 1D migration it would be better to have stamps with smaller $w$, the width of laser beam that was accessible ($10\,\mu$m) gave this limitation for us.

\vspace{0.3cm}

$\blacktriangleright$ \textsc{\textbf{Cell culture and seeding}}

First, cells were cultured at 37$^{\circ}$C and 5$\%$ CO$_2$ atmosphere. Following growth, they were detached using trypsin-EDTA, resuspended in complete culture medium (RPMI + 10 $\%$ FBS + 1 $\%$ penicillin--streptomycin) and finally deposited on the gel.
A small drop ($50\,\mu$L) containing $\sim$ 2,000 cells was set onto the micropatterned gel surface (1.5 cm $\times$ 1.5 cm) bound to the bottom of a Petri dish, and left to adhere for 15 min. Then 2 mL of culture medium were added into the Petri dish. It was then possible to observe them individually under the microscope using the green channel (FITC) of the fluorescence microscope.

\vspace{0.3cm}

$\blacktriangleright$ \textsc{\textbf{live cell microscopy}}

We used an Olympus IX83 inverted microscope equipped with a Hamamatsu Camera (Orca G) to capture fluorescent images in green (FITC), blue (DAPI), and red (TRITC). Thus,  it is possible to capture fluorescence of the actin cytoskeleton using the green channel (FITC), and the beads fluorescence within the substrate via the blue channel (DAPI) of the microscope. TRITC was only used in a few cases to check the track width, using rhodamin fibronectin (Cytoskeleton, Inc.).

The microscope is equipped with the possibility to program the image acquisition mode (CellSens$^{TM}$). The loop we chose allowed to quickly capture two images (green and blue).  It is important to select the appropriate image frequency, since we need to capture fast actin polymerization but we should not expose the cells to too much light (phototoxicity).
Our loop was a series of 15 pairs of images (FITC then DAPI) taken every 5 seconds, then a 15--minute pause was held before starting a new series of images. Thus actin fibers and beads could be followed over short periods of time during cell migration for about 30 minutes.

At the end of the experiment, we added trypsin to the Petri dish to detach the cells from the substrate. After waiting for 15 min, the last beads image was captured to obtain the initial beads position when the gel is in a rest state.
Finally, series of FITC and DAPI images could be treated using image processing in Matlab$^{TM}$ and Fiji$^{TM}$ software for tracking beads and PIV analyses of actin motion.
\vspace{0.3cm}

$\blacktriangleright$ \textsc{\textbf{Image Processing}}

\subsection*{Image acquisition}

In this work, a code formerly written (\cite{Matthieu2019,Amir2020}) was improved. The initial code is divided into four main parts. It reads 16-bit images taken from the microscope and converts them into a data sheet. Next, it finds the cell, produces masks at different times and calculates cell contours. Then, it finds the beads and follows their movement in time. Finally, it selects the beads under the cell and calculates the total displacements of beads (DAPI) between the rest state and the present configuration of the gel.

Due to the addition of trypsin to detach cells and also possible shifts of the camera (thermal changes), it was necessary to rewrite the code in order to process images. Here, the DAPI images (15 DAPI images taken every 5 seconds + DAPI image taken after adding trypsin to the Petri dish) as blue color and FITC images (15 FITC images taken every 5 seconds + blank image corresponding to the absence of cell on the substrate) as green color were merged. Then we used the StackReg plugin \cite{thevenazstackreg} in the Fiji software to shift all images in order to correct for any camera shift. The idea of this plugin is to align or match a stack of images. When the plugin has finished, the current slice works as a global anchor.

Then by separating the blue component from the green one, we could track beads in time using the DAPI images (saved as blue color images) and achieve PIV (Particle Image Velocity) measurements on FITC images (saved as green color images).

\subsection*{Tracking Beads}
\begin{figure}[ht]
    \centering
    \includegraphics[scale = 0.6]{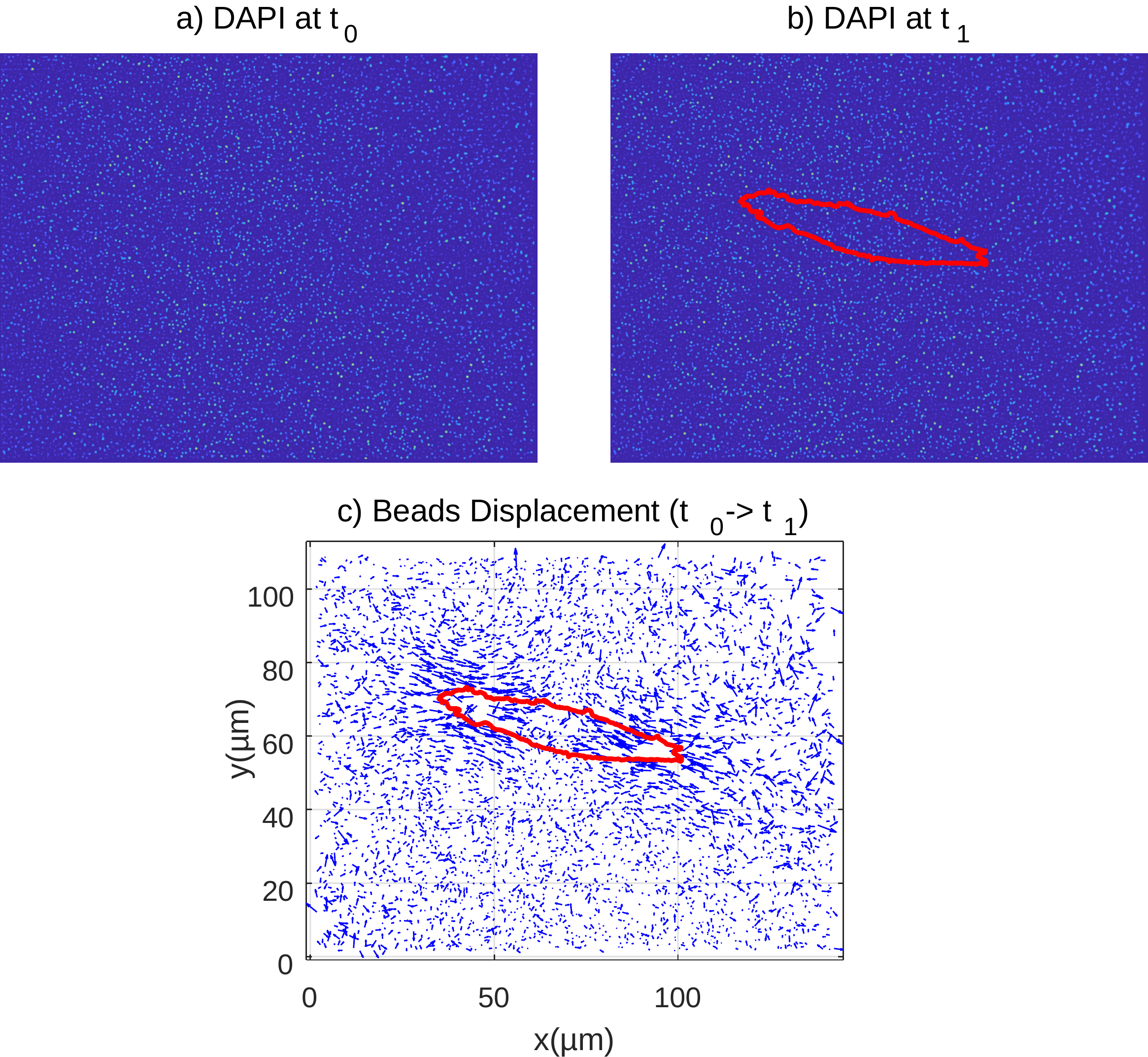}
    \caption{DAPI images. a) time $t_0$ (after adding trypsin), b) time $t_1$; c) Beads displacement between $t_0$ and $t_1$}
    \label{fig:Tracking Beads}
\end{figure}

By using a previously written Matlab code (\cite{Matthieu2019}), we could find bead positions at each time (for example DAPI images at $t_0$ and $t_1$ in Fig. \ref{fig:Tracking Beads}). The particle tracker is a routine adapted for Matlab of a previous program \cite{crocker2011particle}. This function reads the first DAPI image and processes it interactively to figure out what settings are needed in order to identify all the beads. It then repeats this process on all the images. Finally, it links the beads coordinates to form trajectories.

By replacing the last image (obtained after adding trypsin, Fig. \ref{fig:Tracking Beads}a at time $t_0$) from the end of the treated DAPI stack to the first and running the tracking function in Matlab, we obtain a list of trajectories including the beads identities numbered from 1 to the total number of beads as well as their positions ($x$,$y$) at each time $t$ (for example at time $t_1$, Fig. \ref{fig:Tracking Beads}b ). Using this data, it is possible to find displacements for each bead between two times (Fig. \ref{fig:Tracking Beads}c).

\subsection*{Actin and PIV}

\begin{figure}[htb]
    \centering
    \includegraphics[scale = 0.6]{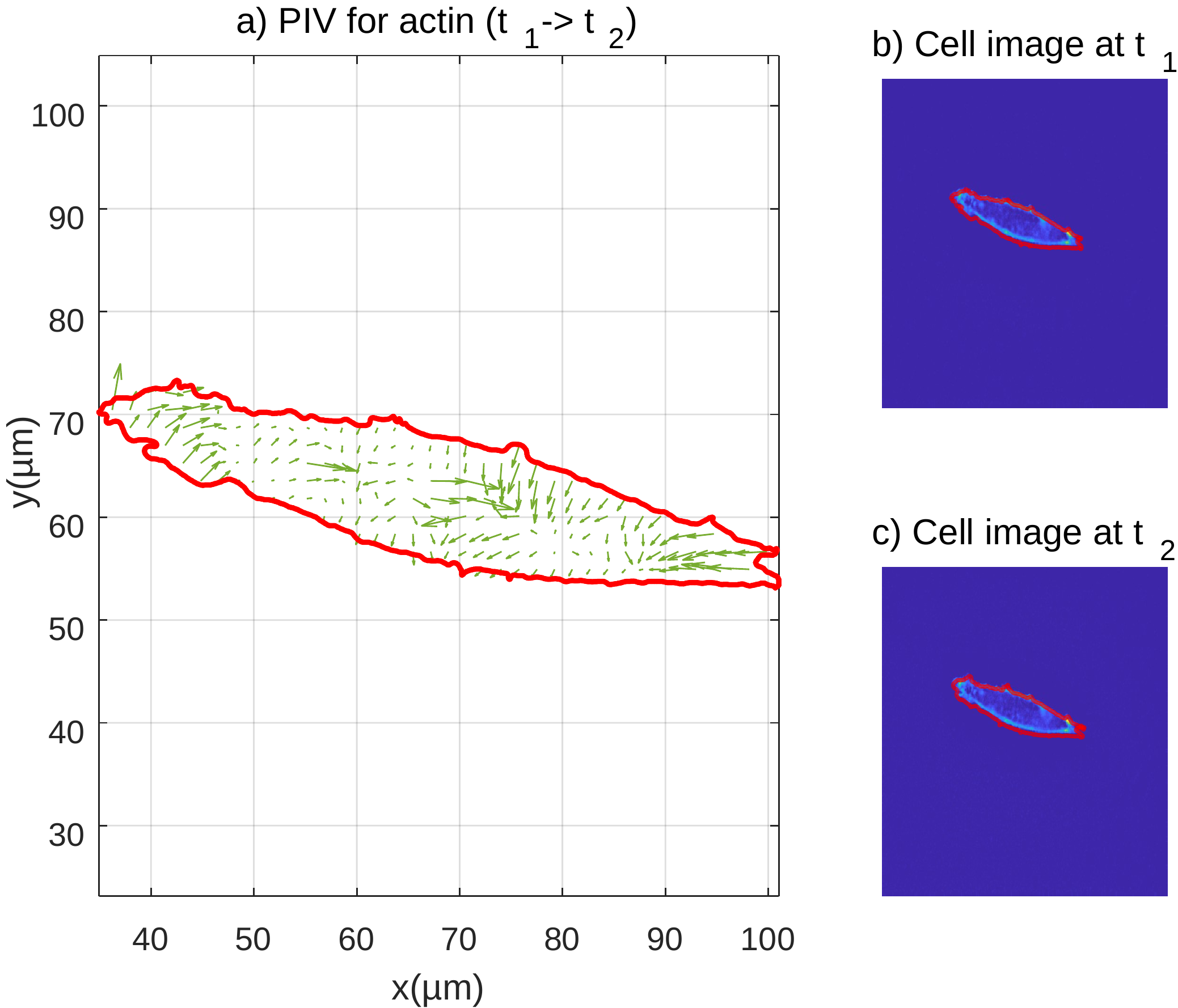}
    \caption{Actin displacements between $t_1$ and $t_2=t_1+5$s (a); Cell shape at $t_1$ (b) and $t_2$ (c)}
    \label{fig:PIV Actin}
\end{figure}

We found Particle Image Velocimetry (PIV) to be a suitable technique to estimate the actin velocity. In general, PIV is an optical technique of flow visualization used to obtain instantaneous velocity measurements and related properties in fluids. Basically, a pair of images is divided into smaller areas named interrogation windows. The cross-correlation between these image sub-regions measures the optical flow (incremental displacement or velocity of the objects) between the two images. To improve resolution, higher PIV resolution can be achieved by progressively decreasing the interrogation window size \cite{tseng2014piv}. The PIV analysis was conducted using the MATLAB tool "PIVlab" \cite{thielicke2014pivlab}, typical outputs from the analysis are shown Fig.~\ref{fig:PIV Actin}.

\begin{figure}[!ht]
    \centering
    \includegraphics[height=.95\textheight]{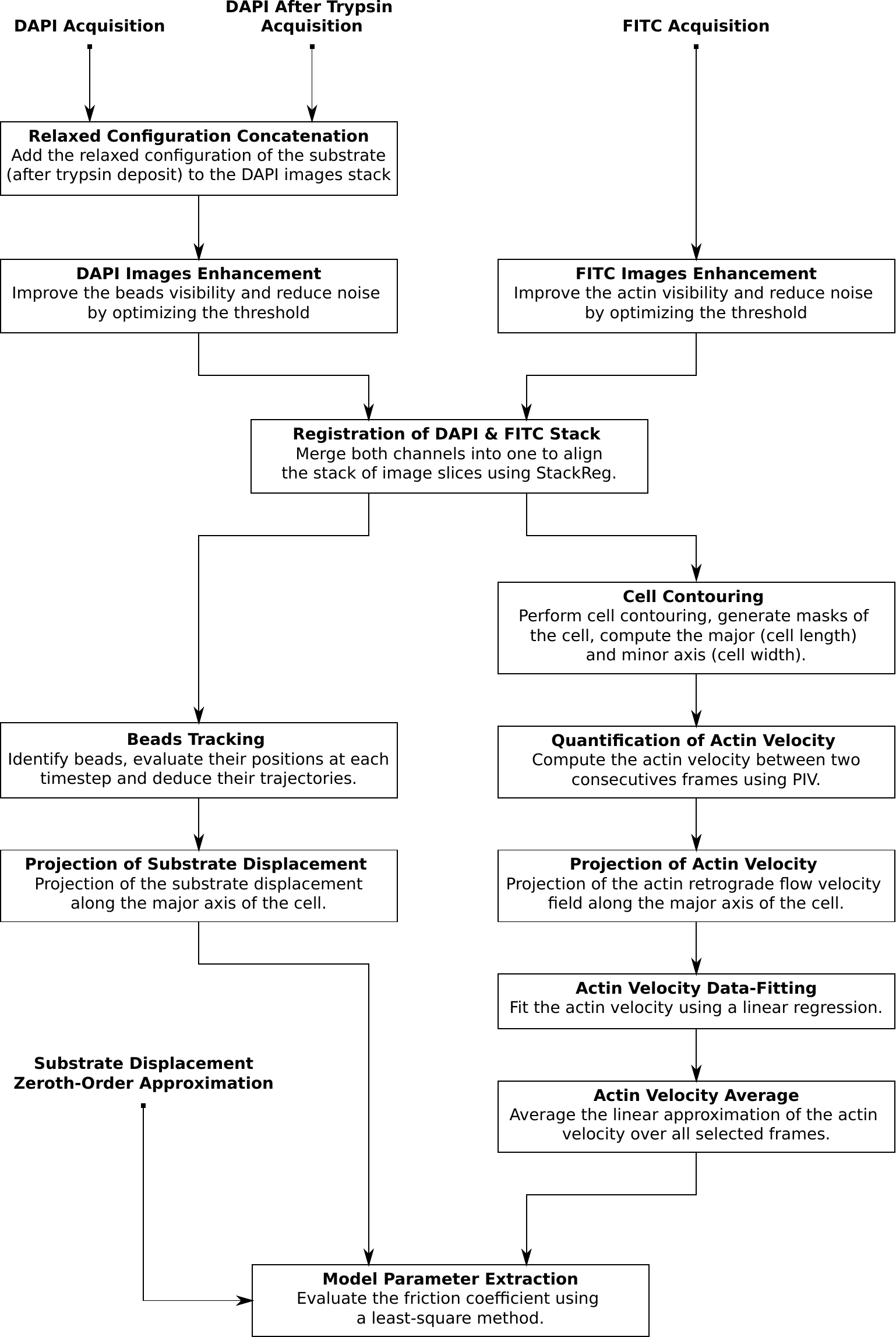}
    \caption{Post-processing method to extract the friction coefficient $\xi$ from the experimental data.}
    \label{fig:postprocessing_diagram}
\end{figure}

The post-processing protocol is summarized in Fig.~\ref{fig:postprocessing_diagram}.
\FloatBarrier

\section{\textcolor{red}{Justification of the semi-infinite approximation for the substrate}}
\label{sec:semi_inf_approx}

\textcolor{red}{To justify the use of our semi-infinite approximation in our experimental context, we set a simple homogeneous loading where $T_x(x,t)=T_0$ and $x\in [-L/2,L/2]$ and compare the ensuing averaging displacement at the surface and at a certain depth in the bulk. Thus with  
$$\bar{\bar{u}}_x(z)=\frac{3T_0}{4\pi \delta L E_s}\int_{-L/2}^{L/2}\int_{-L/2}^{L/2}\int_{-\delta/2}^{\delta/2}\int_{-\delta/2}^{\delta/2}\frac{ \left(2 (x-x')^2+(y-y')^2+z^2\right)}{((x-x')^2+(y-y')^2+z^2)^{3/2}}\mathrm dx'\mathrm dx\mathrm dy'\mathrm dy,$$
we wish to find $z$ such that $\bar{\bar{u}}_x(z)\ll \bar{\bar{u}}_x(0)$. Assuming that both $L\gg \delta$ and $|z|\gg \delta$, we find that at leading order,
$$\bar{\bar{u}}_x(0)=\frac{3\delta T_0}{\pi E_s}\tanh\left( \frac{2L}{\delta}\right)$$
and ($z$ being negative) 
$$\bar{\bar{u}}_x(z)=-\frac{3 L \delta T_0}{4\pi E_s z}.$$
Thus, in order to have $|\bar{\bar{u}}_x(z)/\bar{\bar{u}}_x(0)|\ll 1$, we typically need to impose that $|z|\gg L$. More precisely, in our case, the cell is typically $L\sim 50$ $\mu$m and the substrate thickness is on the order of 70 $\mu$m. Such value is roughly an order of magnitude larger than the track width $\delta$. However, 
$$|\bar{\bar{u}}_x(z)/\bar{\bar{u}}_x(0)|\sim\frac{L}{4z} \left(\tanh\left( \frac{2L}{\delta}\right)\right)^{-1} \sim \frac{L}{4z} = \frac{50}{4\times 70} \sim 0.18$$
is still a bit large and thicker substrates should be used to aim at a quantitative agreement.}

\bibliography{PhD_Haythem}
\bibliographystyle{elsarticle-harv}

\end{document}